\documentclass{article}

\usepackage{emulateapj}

\righthead{DUST IN THE IONIZED MEDIUM OF THE GALAXY}
\lefthead{Howk \& Savage}

\def\s3{S$\;${\small\rm III}\relax}
\def\p3{P$\;${\small\rm III}\relax}
\def\al3{Al$\;${\small\rm III}\relax}
\def\fe3{Fe$\;${\small\rm III}\relax}
\def\si2{Si$\;${\small\rm II}\relax}
\def\si3{Si$\;${\small\rm III}\relax}
\def\zn2{Zn$\;${\small\rm II}\relax}
\def\c2star{C$\;${\small\rm II}$^*$\relax}

\def\HI{H$\;${\small\rm I}\relax}
\def\HII{H$\;${\small\rm II}\relax}
\def\Ha{H$\alpha$\relax}

\def\wave#1{$\lambda$#1 \AA\relax}
\def\twowave#1{$\lambda \lambda$#1 \AA\relax}

\def\kms{km~s$^{-1}$\relax}
\def\etal{{\em et al.}\relax}

\def\percm3{cm$^{-3}$}
\def\percolumn{cm$^{-2}$}

\def\cloudy{CLOUDY\relax}
\def\copernicus{{\em Copernicus}\relax}

\def\wham{WHAM}
\def\hst{{\em HST}}

\def\ghrs{GHRS}
\def\costar{COSTAR}

\def\mucol{$\mu$~Col\relax}
\def\xiper{$\xi$~Per\relax}
\def\betasco{$\beta^1$~Sco\relax}
\def\zoph{$\zeta$~Oph\relax}
\def\rholeo{$\rho$~Leo\relax}
\def\hd18100{HD~18100\relax}
\def\93521{HD~93521\relax}
\def\3c273{3C~273\relax}

\begin{document}

\submitted{Accepted for publication in {\em The Astrophysical
Journal}}

\title{Dust in the Ionized Medium of the Galaxy: \\ GHRS Measurements
	of \ion{Al}{3} and \ion{S}{3}\altaffilmark{1}} 

\altaffiltext{1}{Based on observations made with the NASA/ESA Hubble
	Space Telescope, obtained from the data archive at the Space
	Telescope Science Institute. STScI is operated by the
	Association of Universities for Research in Astronomy,
	Inc. under the NASA contract NAS 5-26555. }

\author{J. Christopher Howk \& Blair D. Savage}
\affil{Department of Astronomy, University of Wisconsin-Madison, 
	Madison, Wi. 53706 \\ Electronic mail:
     howk@astro.wisc.edu, savage@astro.wisc.edu}

\authoremail{howk@uwast.astro.wisc.edu}


\begin{abstract}

We present interstellar absorption line measurements of the ions
\ion{S}{3} and \ion{Al}{3} towards the stars $ \beta^1$~Sco,
$\mu$~Col, $\xi$~Per, $\zeta$~Oph, $\rho$~Leo, and HD~18100 using
archival data from the Goddard High Resolution Spectrograph on board
the {\em Hubble Space Telescope}.  The ions \ion{Al}{3} and \ion{S}{3}
trace heavily depleted and non-depleted elements, respectively, in
ionized gas along the sightlines to these late-O/early-B stars.  We
use the photoionization equilibrium code CLOUDY to derive the
ionization correction relating the ratio $N(\mbox{\ion{Al}{3}}) /
N(\mbox{\ion{S}{3}})$ to the gas-phase abundance [Al/S]$_i \, (\equiv
\log \{N({\rm Al})/N({\rm S}) \}_i - \log \{ {\rm Al/S} \}_\odot$) in
the ionized gas.  For spectral types considered here, the corrections
range from 0.1 to 0.3 dex and are independent of the assumed
ionization parameter, i.e., the ratio of ionizing photon density to
mean electron density.

Using the results of these photoionization models, we find [Al/S]$_i
\, \sim-1.0$ in the ionized gas towards $ \beta^1$~Sco, $\xi$~Per, and
$\zeta$~Oph; along the low-density path towards $\mu$ Col we find
[Al/S]$_i \, \approx -0.8$.  Since S is not depleted onto grains these
values of [Al/S]$_i(\approx [{\rm Al/H}]_i)$ imply that Al-bearing
grains are present in the ionized nebulae around these stars.  If the
WIM of the Galaxy is photoionized by OB stars, the observations of
$\rho$~Leo and HD~18100 imply $[{\rm Al/S}]_i = -0.4$ to $-0.5$ in the
WIM and thus the presence of dust grains containing Al in this
important phase of the ISM.  While photoionization appears to be the
most likely origin of the ionization for \al3\ and \s3, we cannot rule
out confusion from the presence of hot, collisionally ionized gas
along the sightlines to \betasco\ and \hd18100.  We find that
[Al/S]$_i$ in the ionized gas along the six sightlines is
anti-correlated with the electron density and average sightline
neutral density.  The degree of grain destruction in the ionized
medium of the Galaxy is not much higher than in the warm neutral
medium.  The existence of grains in the ionized regions studied here
has important implications for the thermal balance of these regions.

\end{abstract}


\section{INTRODUCTION}

Warm ($10^4$ K) ionized hydrogen is an important component of our
Galaxy's interstellar medium (ISM).  The diffuse warm ionized medium
(WIM) of the Galaxy has a mass surface density one third that of
neutral hydrogen (\HI), with an extended vertical scale height ($h_z
\approx 900$ pc), and a power requirement equivalent to the total
kinetic energy injected into the ISM by supernovae (Kulkarni \& Heiles
1987; Reynolds 1991b).  The WIM has principally been studied through
the detection of faint emission lines (e.g., Reynolds \etal\ 1998b)
and through the dispersion measures of radio emission from distant
pulsars (Taylor \& Cordes 1993; Reynolds 1991a).

The goal of this work is to determine if there is evidence for dust
grains in the ionized medium of the Galaxy, including low-density
\HII\ regions and the diffuse WIM.  The distinction between \HII\
regions and the WIM in this work is mainly one of distance from the
source of ionization and possibly fractional ionization.  The
existence of dust in the WIM of the Galaxy has important ramifications
for the heating (and cooling) of the gas and possibly also for the
power requirements of the ionization.  Photoelectric emission of
electrons from the surfaces of dust grains is an important source of
heating in the warm neutral medium (WNM) of the Galaxy (Wolfire \etal\
1995a).  Reynolds \& Cox (1992) have shown that, if grains are present
in the WIM, photoelectric emission may be the dominant source of
heating in the WIM and may also be responsible for the enhanced
forbidden line strengths that are characteristic of the WIM (e.g.,
Dettmar \& Shulz 1992; Rand 1997, 1998).

To provide evidence for the existence of dust in this important phase
of the ISM, we will measure the gas-phase abundance of Al relative to
S in ionized gas.  In the warm neutral medium relative gas-phase
abundances, derived from absorption line spectroscopy, have been used
to infer the elemental composition of dust grains (see Savage \&
Sembach 1996b and references therein).  We apply this method to the
ionized medium of the Galaxy.

Absorption line spectroscopy of interstellar material yields column
densities of species independent of the prevailing physical conditions
(e.g., $T_e$, $n_e$). In Table \ref{table:uvprobes} we list the
properties of several possibly important probes of weakly ionized gas
in the vacuum ultraviolet.  Along with the wavelengths and $f$-values
(from Morton 1991) for the transitions given, we also give relevant
ionization potentials, logarithmic solar-system abundances of the
elemental species relative to H (Anders \& Grevesse 1989; Grevesse \&
Noels 1993), and representative values of the gas-phase abundances of
each of the elemental species in warm neutral halo gas (see Savage \&
Sembach 1996b).  The lines and ionic species listed in Table
\ref{table:uvprobes} are those likely to provide measurable column
densities in the spectral range accessible to the UV spectrographs
previously or currently on \hst\ ($\sim 1120 - 3200$ \AA) and lines
accessible to the {\em Far Ultraviolet Spectroscopic Explorer}, which
will cover the wavelength range $\,\sim 905 - 1195$ \AA\ after its
launch in early-1999.  The last two columns of Table
\ref{table:uvprobes} give the expected column density of each ionic
tracer for a completely ionized region with halo-like abundances
having $\log N(\mbox{\HII}) = 19.0$ (where $N(\mbox{\HII})$ is
expressed in \percolumn) and total hydrogen density $n_{\rm H} = 1.0$
\percm3, as well as the expected peak optical depth of each line in
the case of pure thermal broadening with $T = 10^4$ K.  These optical
depths will be less if non-thermal broadening plays a significant role
(a likely situation).

Those ions from Table \ref{table:uvprobes} accessible to \hst\ include
\s3, \al3, \p3, \ion{Si}{3}, and \ion{Ti}{3}.  Given the great
strength or weakness of many of the lines from Table
\ref{table:uvprobes}, one is typically able to measure a tracer of a
non-depleted element (see Spitzer \& Fitzpatrick 1993; Fitzpatrick \&
Spitzer 1994, 1997) in the ionized gas (\s3)\footnote{In many
sightlines the S$\,$III \wave{1190.208} line is blended with
intermediate-velocity absorption from the strong Si$\,$II line at
\wave{1190.416} ($v \approx +52$ \kms\ relative to S$\,$III); this is
particularly true for observations made at intermediate resolution
(${\rm FWHM} \gtrsim 15$ \kms).} and a tracer of a refractory element
in ionized gas (\al3) in the \hst\ bandpass.  Relatively little is
known about the Galactic distribution of these ionized gas tracers.
Savage, Edgar, \& Diplas (1990) used the {\em International
Ultraviolet Explorer} to study the distribution of \al3\ in the
Galaxy.  They find an exponential scale-height of $h_{\rm Al\,III} =
1.02^{+0.36}_{-0.24}$ kpc for the distribution of \al3, thereby
showing the distribution of \al3\ is slightly more extended than the
neutral gas ($h_{\rm H\,I} = 0.67^{+0.21}_{-0.16}$ kpc towards these
same objects) and similar to that of free electrons.  From the ratios
of the estimated vertical column densities of \al3\ and free electrons
(which are assumed to trace the protons of the WIM) they infer a
gas-phase abundance of [Al/H]$_i \, > -1.7$ in the ionized gas (using
the solar system abundance Al/H given in Table \ref{table:uvprobes}).
Sembach \& Savage (1992) have also studied the properties of \al3\
absorption towards a number of halo stars.  They find that the
absorption due to \al3\ is significantly different than that of the
more highly-ionized species \ion{Si}{4}, \ion{C}{4}, and \ion{N}{5}.

In this work we present observations of interstellar absorption due to
the ions \s3\ and \al3\ along the sightlines to six late-O/early-B
stars: $\xi$ Persei (HD 24912), $\zeta$ Ophiuchi (HD 149757),
$\beta^1$ Scorpii (HD 144217), $\mu$ Columbae (HD 38666), $\rho$
Leonis (HD 91316), and HD 18100.  These spectra are taken from the
Goddard High Resolution Spectrograph (\ghrs) data archive.  The
stellar and sightline properties relevant to this study are given in
Table \ref{table:stars}.  The first four targets are relatively nearby
disk stars ($z \lesssim 200$ pc).  For these sightlines, the column
density of ionized gas is likely dominated by \HII\ region material,
i.e., by fully-ionized gas immediately surrounding these stars.  For
the latter two stars, which are more distant and at high latitude, the
paths through the ionized ISM are dominated by the WIM, by which we
mean diffuse gas at large distances from any ionizing star.  The
species \al3\ and \s3\ require more energy for creation than the H$^+$
typically used to probe the WIM; however, these ions likely represent
significant fractions of the S and Al associated with the WIM.  In all
cases we depend upon the use of an ionization correction factor to
determine the relative abundances presented here.  However, we will
show that these small corrections are relatively insensitive to model
assumptions.

In \S \ref{sec:data} we discuss the \ghrs\ archival data and our
reductions and analysis of these data.  This includes in \S
\ref{subsec:compfitting} a description of the component-fitting
analysis required for two of our six sightlines, and measurements of
\fe3\ from the literature in \S \ref{subsec:fe3}.  We discuss the
physical conditions and velocity structure of the ionized gas in \S
\ref{sec:velocities}; we include in this a description of the complex
kinematics along the sightline to \zoph\ in \S \ref{subsec:zoph}.  We
use the photoionization code CLOUDY (Ferland 1996; Ferland \etal\
1998) to model ionized regions about hot stars in \S
\ref{sec:abundances} and use the results of these models to derive the
gas-phase Al/S abundance in the ionized gas of the Galaxy.  The
abundances imply the existence of dust in the ionized gas towards
these six stars.  We also discuss in \S \ref{subsec:collisional} the
confusing effects of collisionally ionized gas along our six
sightlines.  In \S \ref{sec:discussion} we discuss the implications of
our derived abundances and dust in the ionized phase of the Galaxy.  A
summary of our work and major conclusions is given in \S
\ref{sec:summary}.

\section{GHRS ARCHIVAL DATA}
\label{sec:data}

In this section we discuss our reduction of the \ghrs\ archival data
and measurements of interstellar absorption features in these data.  A
more complete discussion can be found in Howk, Savage, \& Fabian
(1998).  Included in this section (\S \ref{subsec:compfitting}) is a
description of our approach to model component fitting.  Component
fitting is required to disentangle the \s3\ \wave{1190.208} absorption
line from the strong \ion{Si}{2}\ line at \wave{1190.416} in
intermediate resolution data.  Also, we discuss the literature values
of \fe3\ \wave{1122.526} absorption used in this paper (\S
\ref{subsec:fe3}).

\subsection{Reductions and Measurements}
We have retrieved the \ghrs\ archival data containing the \s3\ and
\al3\ transitions for the stars listed in Table \ref{table:stars}.
The STScI archive identification codes and relevant information
including exposure times, grating mode, and aperture are given for
each spectrum in Table \ref{table:log}.  The spectra have been
calibrated using the standard {\tt CALHRS} routine which includes
conversion of raw counts to count rates and corrections for particle
radiation contamination, dark counts, known diode non-uniformities,
paired pulse events and scattered light.  The final data reduction was
performed using software developed and tested at the University of
Wisconsin-Madison.  This includes the merging of individual spectra
and allowing for additional refinements to the scattered light
correction for the echelle-mode data (Cardelli, Ebbets, \& Savage
1993).

The spectra used for this study include data taken both before and
after the installation of the Corrective Optics Space Telescope Axial
Replacement (\costar). For information regarding the pre-\costar\
performance of the \ghrs, see Heap \etal\ (1995); the post-\costar\
performance of the \ghrs\ is discussed by Robinson \etal\ (1998).  The
pre-COSTAR data used here were taken through the Small Science
Aperture (SSA), thereby avoiding the complication of the very broad
wings found in the pre-COSTAR Large Science Aperture (LSA) line-spread
function (LSF).  The \ghrs\ echelle-mode spectra (Ech-A and Ech-B)
used here have a resolution of $\sim$3.5 \kms\ (FWHM).  The
observations of the \s3\ \wave{1190} region using the G160M grating
have a velocity resolution of $\sim 20$ \kms, while in the \al3\
\twowave{1855, 1862} region the resolution is $\sim 11$ \kms\ (FWHM).
For the G160M data we have used the SPYBAL observations taken before
the science integrations to refine the absolute velocity scales (see
Soderblom, Sherbert, \& Hulbert 1993, 1994).  The SPYBAL observations
contain emission lines observed in the wavelength range
$\sim1500-1540$ \AA\ range.  We have calculated the average offset
between the observed and expected wavelengths of these lines, and
applied these wavelength offsets to the appropriate science
observations (Soderblom \etal\ 1994).  Using the SPYBAL observations,
we find offsets of $-0.003$ \AA\ and $-0.033$ \AA, respectively, are
appropriate for the \s3\ and \al3\ observations towards \hd18100.
Towards \betasco\ the G160M observations of \s3\ require an offset of
$-0.014$ \AA.  These wavelength adjustments should make the absolute
velocities for the G160M observations good to $\sim3$ \kms.  The
velocity scale of the echelle data presented here is likely good to
$\pm$1 resolution element (i.e., 3.5 \kms).

Most of the data presented here employed the FP-SPLIT=4 procedure,
which takes four exposures at different grating carousel positions.
These scans are identified in Table \ref{table:log}.  The FP-SPLIT
exposures can be used to solve for the fixed-pattern noise spectrum of
the \ghrs\ (Cardelli \& Ebbets 1994) and remove it from the
observations.  In general we have solved for the fixed-pattern noise,
but only applied the solution in the cases where the noise spectrum
showed significant structure within a few hundred \kms\ of the line of
interest.  In this way we avoid adding noise to those spectra with few
or no fixed-pattern features.  Only for the \zoph\ scans and the
echelle observations of \al3\ towards \betasco\ was there structure
deemed to be possibly significant to the current work; these are the
only observations for which we have applied the fixed-pattern noise
solutions.

\begin{figure}
\epsscale{0.87}
\plotone{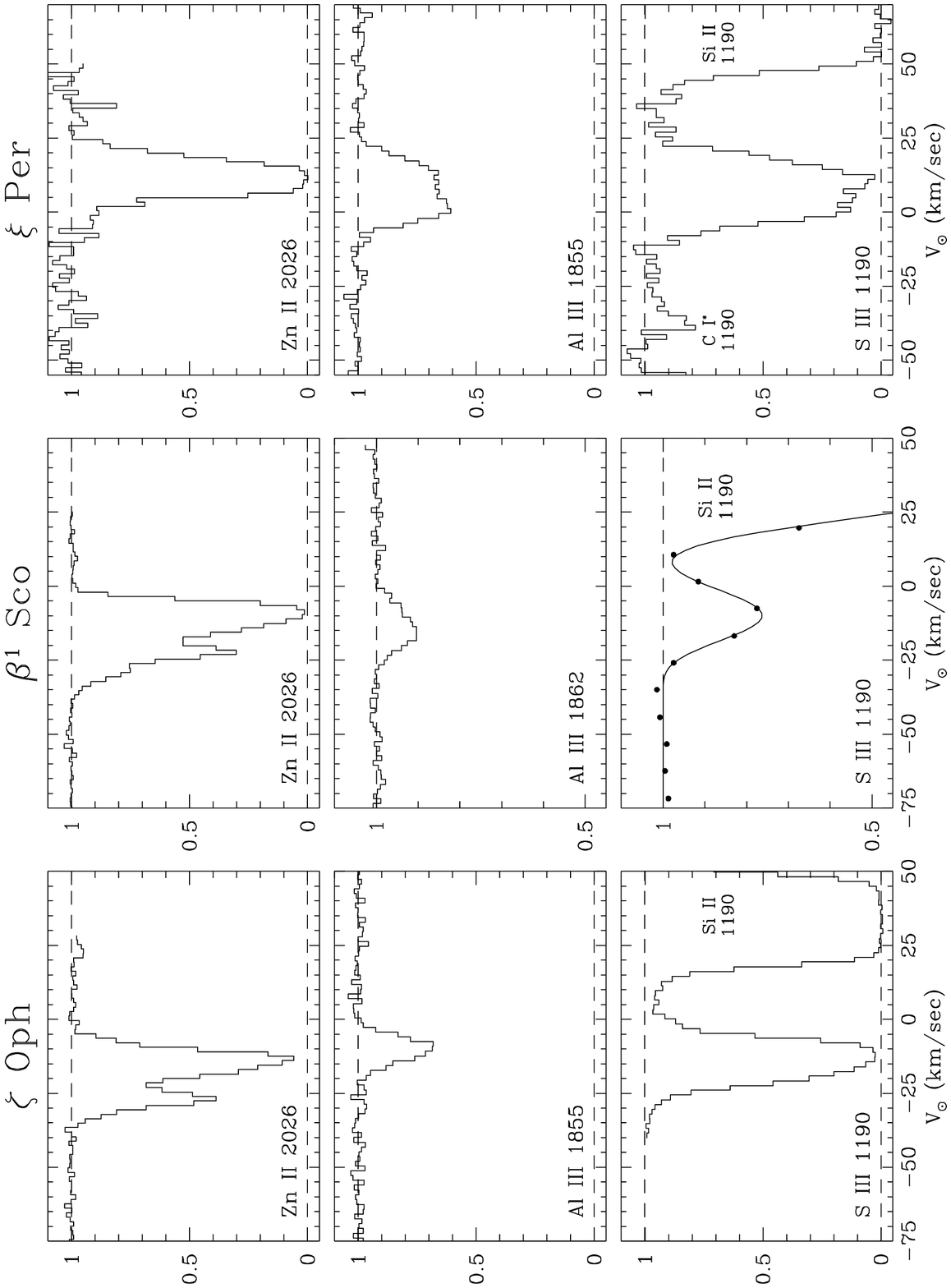} 
\figcaption{Continuum-normalized absorption line profiles of
\protect\ion{Zn}{2}, \protect\al3, and \protect\s3\ for the disk stars
\protect\xiper, \protect\zoph, and \protect\betasco\ are displayed on
a heliocentric velocity scale.  Echelle-mode data are plotted as
histograms.  The G160M data for the \protect\s3\ profile towards
\protect\betasco\ are plotted as points, while the component model
convolved with the instrumental spread function is overplotted as the
solid line (see \S \protect\ref{subsec:compfitting}). Strong
\protect\ion{Si}{2} \protect\wave{1190.416} absorption is seen in the
\protect\s3 \protect\wave{1190.208} region of the spectrum ($v =
+52.4$ \protect\kms\ relative to \protect
\s3). \label{fig:spectra}}
\end{figure}

We have normalized the spectra using low-order ($\leq4$) Legendre
polynomial fits to the local stellar continuum in regions free from
interstellar absorption.  Figures \ref{fig:spectra} and
\ref{fig:spectra2} present the normalized absorption line profiles of
\s3\ and \al3 for the stars considered here.  Also shown in Figure
\ref{fig:spectra} are the profiles of \ion{Zn}{2} \wave{2026.14} or
\wave{2062.66}.  The \ion{Zn}{2} profiles trace the column density of
neutral material along these sightlines.  Absorption from the
\ion{Si}{2}\ \wave{1190} line can be seen in all of the plots showing
the \s3\ absorption.  The expected shift of the \ion{Si}{2}\ line
relative to \s3\ is $+52.4$ \kms.  If intermediate negative velocity
\ion{Si}{2} is present, it can be blended with the \s3 line.  The
\s3+\ion{Si}{2}\ profile towards the star \rholeo\ is a good example
of this behavior; narrow intermediate velocity \ion{Si}{2}\ features
are present at $v_\odot \approx +12$ and $+25$ \kms\ relative to the
restframe \s3\ along this sightline.  These two features make it
impossible to determine the \s3\ absorption for $v_\odot > 5$ \kms.
Therefore our study of the \rholeo\ sightline is limited to $v_\odot <
5$ \kms.  The behavior of \ion{Si}{2} can be determined from the
\ion{Si}{2} \wave{1193.29} transition.

Table \ref{table:columns} contains the measured equivalent widths and
column densities of interstellar \s3\ and \al3\ towards our six
targets, along with the $1\sigma$ uncertainties in these quantities.
These uncertainties include contributions from photon statistics,
continuum placement uncertainties, and zero-level uncertainties.  We
have adopted a 2\% zero-level uncertainty throughout.  This may
overestimate the errors in regions near heavily saturated lines, i.e.,
in the wavelength region of \s3.  See Sembach \& Savage (1992; their
Appendix) and Howk \etal\ (1998) for a more thorough discussion of
these sources of uncertainty.

For the echelle data the values of the column densities presented in
Table \ref{table:columns} were derived by a straight integration of
the apparent column density profiles (Savage \& Sembach 1991).  In the
absence of unresolved saturated structure, $N_a(X^i) = N(X^i)$, where
$N_a(X^i)$ and $N(X^i)$ are the apparent and true column densities of
the ion $X^i$, respectively.  The \s3\ lines observed towards \xiper\
and \zoph\ are the only profiles for which we might expect the
presence of unresolved saturated structure.  Though the profiles seem
resolved by the echelle-mode resolution, the value
$N(\mbox{\ion{S}{3}})$ for the \xiper\ and \zoph\ sightlines may
underestimate the total column density.  For gas at $T=10^4$ K, the
thermal Doppler spread parameters for \s3\ and \al3\ are
$b(\mbox{\s3}) \approx 2.3$ \kms\ and $b(\mbox{\al3}) \approx 2.5$
\kms\ (FWHM$\, \sim 3.8$ and $4.2$ \kms, respectively).  Thus the
profiles of photoionized gas traced by these species should be
marginally resolved, even without the likely addition of non-thermal
broadening.

As mentioned above, intermediate velocity \ion{Si}{2}\ absorption is
seen towards \rholeo\ at velocities near the restframe of \s3.  In
both the \al3\ and \ion{Zn}{2} profiles, an absorbing component is
present near $v_\odot \approx +20$ \kms.  The corresponding \s3\
absorption is blended with the intermediate velocity \ion{Si}{2}\
absorption.  Therefore we report the column densities of \al3\ and
\s3\ for this sightline integrated over the velocity range $v_\odot =
-19$ to $+2$ \kms.  This only includes the lower-velocity \al3\
absorbing component.  We do not expect any \ion{Si}{2}\ absorption in
this velocity range given the profile of the \ion{Si}{2}\
\wave{1193.29} line present in the same \ghrs\ observation.

\subsection{Component Fitting Measurements}
\label{subsec:compfitting}

The \s3\ \wave{1190.208} lines observed towards \betasco\ and
\hd18100\ with the G160M grating are blended with the nearby
\ion{Si}{2} \wave{1190.416} line ($v = +52.4$ \kms\ relative to \s3).
To assess the extent of the blending, we have performed a component
fitting analysis of these profiles using software kindly provided to
us by E. Fitzpatrick and described in Spitzer \& Fitzpatrick (1993;
hereafter SF93).  A model of the interstellar absorption was convolved
with the instrumental LSF appropriate for the pre-COSTAR SSA, and the
value of $\chi^2$ minimized between this model and the data.  The LSF
adopted here for the SSA follows Spitzer \& Fitzpatrick (1995) in
using instrumental LSF given for the G160M grating in the \ghrs\
Instrument Handbook (v3.0; Duncan 1992, Table 4-8).  The LSF is well
characterized by a Gaussian with a FWHM $\approx \, 1.08$ diodes at
the detector array.  From these models, we derive the best-fit column
density, $N_k$, Doppler parameter, $b_k$, and central velocity,
$\langle v \rangle_k$, for the $k$ absorbing components of
\ion{Si}{2}\ and \s3\ along these lines of sight.

For the \betasco\ observations, we have used the the high-resolution
\ghrs\ Ech-B observations of the \ion{Si}{2} \wave{1808.013} line
along this sightline to provide more information about the component
structure of the \ion{Si}{2} \wave{1190} line.  We have fit the
\wave{1808} line with a four-component model and subsequently fixed
the values of $N_k$ and $b_k$ from this fit, though allowing common
shifts in the velocities, to describe the \ion{Si}{2} \wave{1190}
absorption.  The \s3\ column density presented for \betasco\ in Table
\ref{table:columns} is derived from the best-fit one component model.
Figure \ref{fig:spectra} displays the G160M data for this sightline as
points, while the model derived from our fitting is displayed as a
solid line.  The central velocity for the \s3\ profile is $\langle
v_\odot \rangle = -10.0\pm0.4$ \kms\ (fitting error only).  The
$b$-value is not well constrained but is likely significantly less
than the resolution of the G160M observations.

\begin{figure}
\epsscale{0.87} 
\plotone{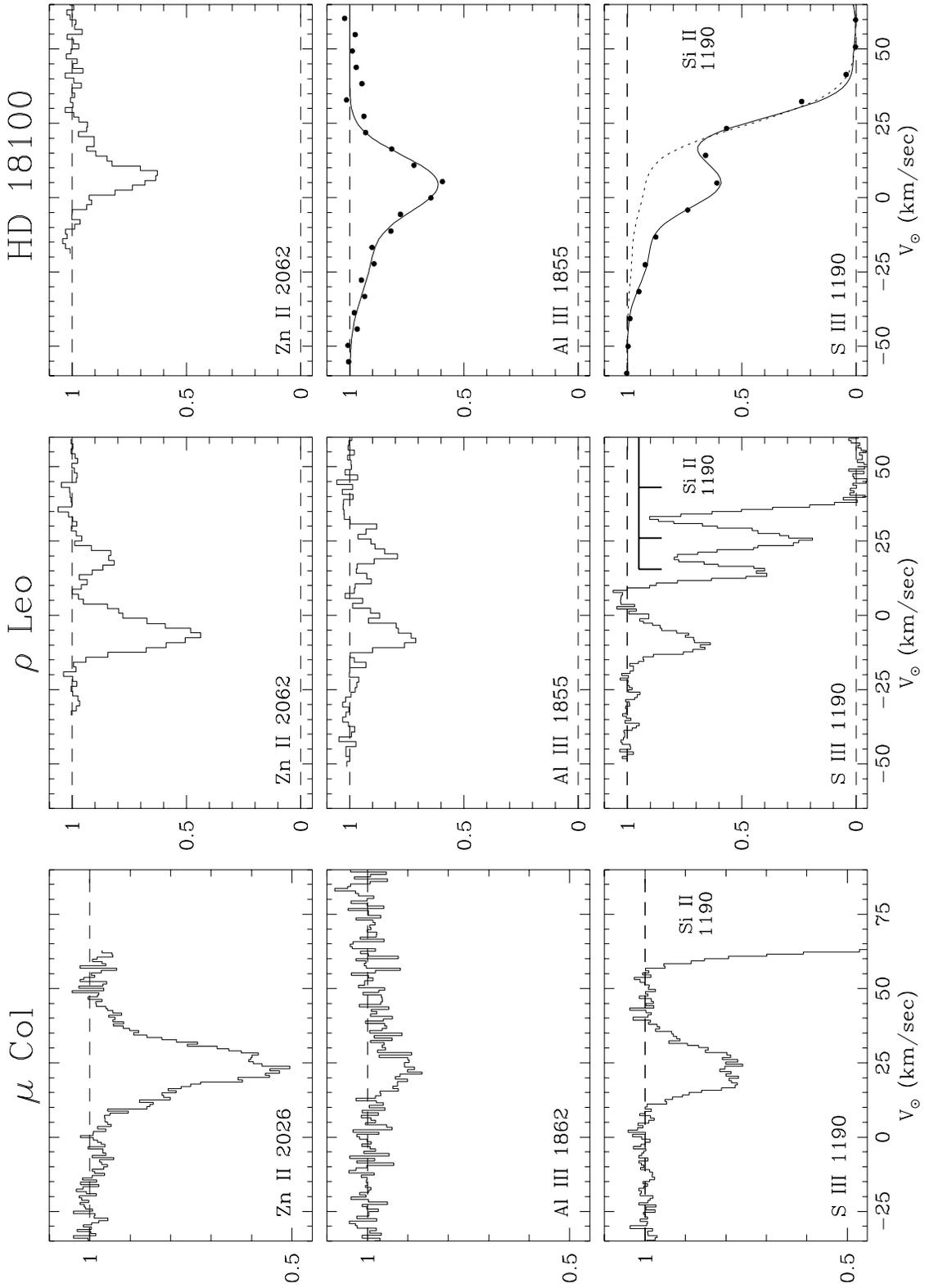}
\figcaption{As for Figure \protect\ref{fig:spectra} but for the
higher-$z$ stars \protect\mucol, \protect\rholeo, and
\protect\hd18100.  The G160M data for the \protect\al3\ and
\protect\s3\ profiles towards \protect\hd18100\ are overplotted with
the component model convolved with the instrumental LSF; the dashed
line in the \protect\s3\ profile for this star shows the absorption
model for \protect\ion{Si}{2} \protect\wave{1190} derived from the
\protect\ion{Si}{2} \protect\wave{1193} transition (see \S
\protect\ref{subsec:compfitting}).  \label{fig:spectra2}}
\end{figure}

For the sightline towards \hd18100, no high-resolution observations of
\ion{Si}{2} were available.  However, we have fit the \ion{Si}{2}
\wave{1193.290} line present in the G160M observation to assess the
contribution of intermediate-velocity \ion{Si}{2} absorption to the
\s3\ profile.  The model fit to the \ion{Si}{2} \wave{1193} profile is
over-plotted as the dotted line on the \s3\ profile for \hd18100\ in
Figure \ref{fig:spectra2}.  This model has been scaled by the
appropriate factor to account for the differing oscillator strengths
of the 1190 and 1193 transitions of \ion{Si}{2}\ and shifted to the
velocity scale of \s3.  It is clear from this presentation that there
is a component of \ion{Si}{2}\ overlapping the \s3\ absorption in
velocity.  We have held the column density of this \ion{Si}{2}
component fixed when fitting the \s3+\ion{Si}{2} blend at 1190 \AA.
We find the \ion{Si}{2} component is relatively weak, of order 12\% of
the \s3\ column density at these velocities.  Table
\ref{table:columns} gives the total column density derived for \s3\
along this sightline.  The total column density of \s3\ we obtain
agrees with the value $\log N(\mbox{\s3}) = 14.26\pm0.06$ derived by
Savage \& Sembach (1996a, hereafter SS96).

We have similarly fit the \al3\ profile for \hd18100\ with a component
model.  We find evidence for at least two components in the \s3\ and
\al3\ absorption along this sightline.  While the $b$-values are
ill-constrained and likely much less than the G160M instrumental
resolution, the column densities and central velocities for the
individual absorbing components of \s3\ are as follows:
\begin{displaymath} 
\langle v_\odot \rangle_1 =   -20\pm6\phn \phn \ \mbox{\kms ;} \ 
\log N(\mbox{\s3})_1      =   13.51\pm0.25; \ \ 
\end{displaymath}
\begin{displaymath} 
\langle v_\odot \rangle_2 =   +5.8\pm0.9  \ \mbox{\kms ;} \   
\log N(\mbox{\s3})_2      =   14.21\pm0.03. \ 
\end{displaymath}
For \al3\ we find:
\begin{displaymath}   
\langle v_\odot \rangle_1 =   -19\pm13 \phn\ \mbox{\kms ;} \ 
\log N(\mbox{\al3})_1     =   12.1 \pm0.4;  \phn \phn
\end{displaymath}
\begin{displaymath} 
\langle v_\odot \rangle_2 =   +4.8\pm1.6  \ \mbox{\kms ;} \   
\log N(\mbox{\al3})_2     =   12.58\pm0.11
\end{displaymath}
We do not believe unresolved saturation is a significant problem for
either species for this sightline.  Column densities derived for the
two \al3\ transitions through a straight integration of the are the
$N_a (v)$ profiles are the same within the errors, and their $N_a (v)$
profiles show no evidence for saturation effects (see Savage \&
Sembach 1994; their Figure 4).  For the \s3\ profile we find that the
equivalent width is within the range covered by the two \al3\
transitions, and the peak optical depth is very similar to that found
in the \al3\ \wave{1855} profile (though it includes a small
contribution from the overlapping \ion{Si}{2} \wave{1190} line).  The
data for \al3\ and \s3\ along the sightline to \hd18100\ are shown in
Figure \ref{fig:spectra2} as points, while the component models for
each species are plotted as solid lines.  Considering the difficulty
in deriving precise values for the $\langle v_\odot \rangle_1 \sim
-20$ \kms\ component, we will use the integrated sight-line column
densities in our analysis.  We also note that the component structure
buried in the intermediate resolution G160M data may be significantly
more complex than our component-fitting analysis suggests (see SS96;
Ryans, Sembach, \& Keenan 1996).

\subsection{Literature Measurements of Fe$\,${\small III}}
\label{subsec:fe3}

For the stars \mucol, \betasco, and \zoph\ measurements of the \fe3\
\wave{1122.526} line are available.  Howk \etal\ (1998) report on the
profile and column density of \fe3\ towards \mucol\ as measured with
the \ghrs\ using the G140M grating.  Savage \& Bohlin (1979) and
Morton (1975) report on the gas-phase column density of \fe3\ towards
\betasco\ and \zoph, respectively, based upon \copernicus\
measurements.  The column densities derived by these authors for \fe3
are given in Table \ref{table:columns}.  The data derived from the
earlier \copernicus\ papers have been modified to reflect more recent
determinations of the $f$-values of this transition (Morton 1991).  We
adopt $f = 0.0788$, which amounts to a modification of $-0.15$ dex to
the column densities adopted in the works of Savage \& Bohlin (1979)
and Morton (1975).

The \fe3\ absorption towards \betasco\ and \zoph\ could be
substantially blended with \ion{C}{1} absorption at \wave{1122.438}
and \wave{1122.518}.  Howk \etal\ (1998) have put restrictive limits
on the degree of contamination to the \fe3\ absorption towards \mucol.
They find the level of contamination is not significant compared with
the errors.  For the \copernicus\ results, we have estimated the
errors quoted in Table \ref{table:columns}, though these column
densities should be considered upper limits due to the possible
contamination from \ion{C}{1} absorption.

\section{PHYSICAL CONDITIONS AND VELOCITY STRUCTURE OF THE IONIZED GAS}
\label{sec:velocities}

The sightlines to the stars in our sample cover a wide range of
physical conditions and environments.  The absorption lines due to
ionized gas along the sightlines towards \betasco, \zoph, and \xiper\
are likely dominated by photoionized gas in the \HII\ regions of these
disk stars.  The sightlines towards the higher-latitude stars sample
the WIM of the Galaxy.  Towards \mucol, the gas being sampled may well
be a mixture of both \HII\ region and WIM gas, though to our knowledge
no \HII\ region has previously been identified about this star.  Given
the range of conditions and velocity structure, we mention some of the
more important aspects of these ionized gas properties here.

\subsection{H$\,${\small II} Region Sightlines}
\label{subsec:hiiregions}

The \HII\ regions of some of our disk stars have been studied in
detail.  Reynolds \& Ogden (1982) and Reynolds (1988b) have studied
the \HII\ region S27 surrounding \zoph.  Their analysis of faint \Ha\
and [\ion{S}{2}] emission from the nebula suggests $\langle n_e^2
\rangle^{1/2} = n_e f^{1/2} \approx 4$ \percm3\ with $T_e \approx
6700$ K, where $n_e$ is the local electron density and $f$ is the
volume filling factor.  Reynolds (1988b) has similarly studied Sivan
4, the \HII\ region surrounding \xiper.  The \Ha\ and [\ion{S}{2}]
emission suggest $\langle n_e^2 \rangle^{1/2} = n_e f^{1/2} \approx
1.4$ \percm3\ for this nebula with $T_e \approx 8000$ K.  Howk \etal\
(1998) and Shull \& York (1977) have derived $\langle n_e \rangle =
0.2$ \percm3\ for the ionized gas along the line of sight towards
\mucol, which is similar to that in the Galactic WIM.  This average
density has been derived from analyses of the excited states of
\ion{Si}{2} and \ion{N}{2}, respectively, and is a lower limit to the
true electron density.

The \Ha\ emission towards \xiper\ is centered at $v_\odot = +7$ \kms\
(Reynolds 1988b).  Emission from the nebula S220 1$^\circ$ to the
north, which is also thought to be powered by \xiper, is centered at
$v_\odot = +12$ \kms.  These velocities are well within the limits of
\al3\ and \s3\ absorption towards this star.  Towards \zoph\ the \Ha\
emission is centered at $v_\odot = -13$ \kms\ (Reynolds 1988b), which
is roughly consistent with the \al3\ absorption and the
positive-velocity edge of the \s3\ absorption.  

Recent observations with the Wisconsin \Ha\ Mapper (\wham) Fabry-Perot
instrument (M. Haffner, private communication) have similarly shown
that \Ha-emitting gas towards \mucol\ lies at velocities quite similar
to the center of \s3\ and \al3\ absorption along this sightline.  The
sightline towards this star may be associated with an \HII\ region,
the WIM along this sightline, or a combination of both.  At the
distance and latitude of \mucol, we expect only $\sim20\%$ of the WIM
in the direction of \mucol\ to be in front of the star (Reynolds
1991b).  Given that the \Ha\ emission in this direction is at
velocities consistent with the \s3\ profile presented here, even
though the WIM along this direction should mostly come from beyond the
star, we will assume the emission and absorption along the sightline
towards \mucol\ are mostly probing \HII\ region gas.

When comparing the velocities of the tracers of ionized gas being
studied here and tracers of neutral gas, such as \zn2, we see three
distinct arrangements in the nearby disk stars.  The ionized gas along
the sightline to \betasco\ is found at velocities between those of the
major components observed in the low-ionization species.  The \al3\
and \s3\ absorption towards \mucol\ resides at velocities similar to
the principal component seen in neutral gas tracers.  For each of the
stars \zoph\ and \xiper\ we find a component of low-ionization
absorption that is aligned with one edge of the \s3\ profile.  The
low-ionization components found at this edge are significantly
narrower than the \s3\ absorption profile.  There is no systematic
trend when comparing the velocity structure of neutral and ionized gas
tracers along these low-$z$ sightlines.

\subsubsection{The Complicated Case of \zoph}
\label{subsec:zoph}

The velocity structure observed in tracers of ionized gas along the
path to \zoph\ is quite complex and warrants a more complete
description.  The profiles of \al3, \s3, and \zn2\ towards \zoph\
shown in Figure \ref{fig:spectra} have distinctly different velocity
structure.  The \s3\ absorption is broader than either the \zn2\ or
\al3\ profiles and is asymmetric.  The \zn2\ absorption lines up well
with the negative velocity edge of the \s3\ profile, while the \al3\
absorption is aligned quite well with the positive velocity edge.

\begin{figure}
\plotone{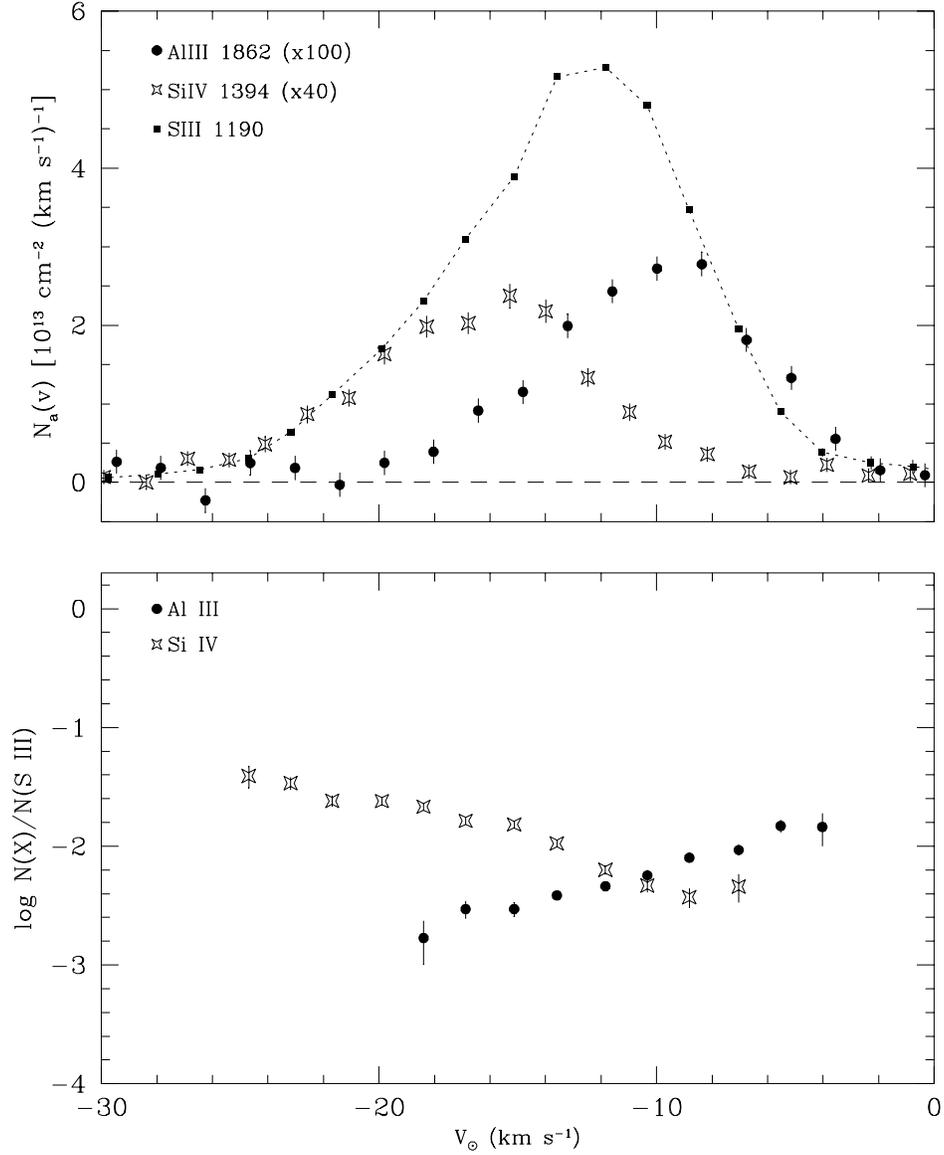} \figcaption{The top panel shows the apparent
column density profiles of \protect\al3, \protect\s3, and
\protect\ion{Si}{4} observed towards \protect\zoph\ at 3.5 \protect\kms\
resolution displayed on a heliocentric velocity scale.  The profiles
of \protect\al3\ and \protect\ion{Si}{4} have been scaled upwards by
factors of 100 and 40, respectively.  The bottom panel shows the
logarithm of the ratio of \protect\al3\ and \protect\ion{Si}{4} to
\protect\s3\ as a function of velocity.  \label{fig:zoph}}
\end{figure}

Sembach, Savage, \& Jenkins (1994) studied the \al3, \ion{Si}{4}, and
\ion{C}{4} absorption along this sightline in some detail.  These
authors attributed the \al3\ and \ion{Si}{4} absorption to the
photoionized nebula about \zoph\ and suggested the velocity offsets
seen between these two species indicated the \HII\ region was
expanding.  In Figure \ref{fig:zoph} we plot the apparent column
density profiles (see Savage \& Sembach 1991) of the ions \al3,
\ion{Si}{4}, and \s3.  The \ion{Si}{4} and \al3\ profiles have been
scaled upwards by factors of 40 and 100, respectively.  The lower
panel of Figure \ref{fig:zoph} shows the log of the apparent column
density ratios of \al3\ and \ion{Si}{4} to \s3\ as a function of
velocity.  The value of $\log [N(\mbox{\al3}) / N(\mbox{\s3}) ]$
varies by about one dex between the peak of the \al3\ profile at
$v_\odot \approx -8$ \kms\ and the peak \ion{Si}{4} absorption near
$v_\odot \approx -15$ \kms.  The component at $v_\odot \approx -15$
\kms\ coincides in velocity with the dense neutral/molecular cloud
seen in the \zn2\ profile (see also Savage, Cardelli, \& Sofia 1992).

Figure \ref{fig:zoph} shows the complexities that arise when looking
in detail at a given interstellar sightline.  The other sightlines in
our sample may be similarly complex, with complications hidden in the
absorption profiles.  The modelling we present in \S
\ref{subsec:cloudy} is highly idealized and does not account for the
physics of stellar wind bow-shocks, for example.  For the \zoph\
sightline we will give results derived from the integrated sightline
values of $\log N(\mbox{\al3})$ and $N(\mbox{\s3})$; we will, however,
note the results for the case of $\log N(\mbox{\al3}) / N(\mbox{\s3})
\approx -2.0$ along this sightline, which is appropriate for gas
associated with the strongest component of \al3\ along this sightline.

\subsection{The High-Latitude Sightlines to \hd18100\ and \rholeo}
\label{subsec:hilatitude}

The average electron densities towards the higher-latitude stars can
be estimated using the $^2P_{3/2}$ fine structure level of \ion{C}{2}
(denoted \c2star).  Savage \& Sembach (SS96) have studied the
excitation of \c2star\ towards \hd18100.  They find, assuming the
excitation is caused by electron collisions, $\langle n_e \rangle =
0.071$ \percm3.  This treatment assumes that all of the \c2star\ and
\ion{S}{2} along the line of sight arise in the same gas and is a
lower limit to the true electron density in the ionized gas.  The
value $\langle n_e \rangle$ for \rholeo\ given in Table
\ref{table:stars} is our own determination assuming electron
collisions are populating the upper fine-structure level of
\ion{C}{2}.  We have measured $\log N(\mbox{\c2star}) = 14.13\pm0.02$
and $\log N({\rm S\, II}) = 15.51\pm0.02$ from archival \ghrs\ data
for this sightline.  The column of \c2star\ was derived from
observations using the G160M grating and should be considered a lower
limit given the possible presence of unresolved saturated structure.
The average electron density is then determined using Eqn. (7) of
SF93, though we adopt [C/S]$\, \approx \, $[C/H]$\, = -0.4$ from
Cardelli \etal\ (1996).  We find $\langle n_e \rangle \geq
(0.074\pm0.005) \, (T/6000\, {\rm K})^{0.5}$ \percm3.  Thus, $\langle
n_e \rangle$ towards \rholeo\ is similar to that towards the halo
stars \hd18100\ and HD~93521 as well as the direction towards 3C~273,
which also probes halo gas.\footnote{The cooling rate per neutral
H-atom for the \rholeo\ sightline in the C$\,$II 158 $\mu$m line is
therefore also similar to these other halo sightlines, with $l_c
\gtrsim (1.4\pm0.3)\times 10^{-26}$ ergs s$^{-1}$ H-atom$^{-1}$.  The
cooling in the 158 $\mu$m line along the \zoph\ and \xiper\ sightlines
may be as low as a factor of five below this value (Gry, Lequeux, \&
Boulanger 1992), further illustrating the differences in the ionized
gas along the low- and high-latitude sightlines.}  Reynolds (1991a)
finds the high-latitude WIM is clumped in regions having electron
densities $\langle n_e \rangle \approx 0.08$ \percm3, a value quite
close to the average densities derived for the high-latitude
sightlines discussed here.

For the two distant high-latitude stars \hd18100\ and \rholeo\ we find
a very good velocity correspondence between low-ionization gas (traced
for example by \zn2) and ionized gas traced by \al3\ and \s3.  Savage
\& Sembach (1994, 1996) present GHRS observations of the high and low
stages of ionization, respectively, towards \hd18100.  The
low-ionization lines along this sightline are found at the same
velocities as \al3\ and \s3, as well as the high-ionization species
such as \ion{Si}{4} and \ion{C}{4}; Savage \& Sembach (1994) find the
observed profile widths increase with the ionization potential of the
species.

\begin{figure}
\plotone{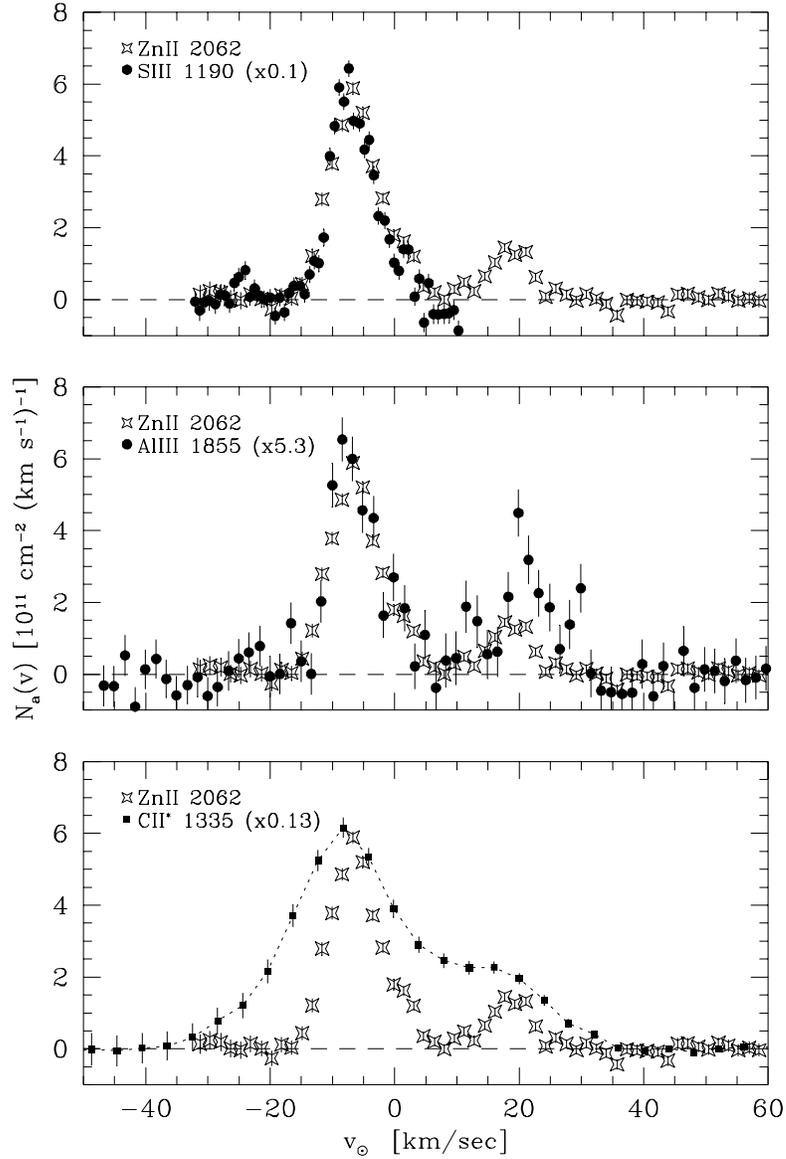} 
\figcaption{Apparent column density profiles of \protect\s3,
\protect\al3, \protect\c2star, and \protect\zn2\ towards
\protect\rholeo\ are plotted against heliocentric velocity.  The
\protect\c2star\ observations were obtained with the G160M grating,
the other measurements were made with the \protect\ghrs\ echelle
gratings.  The profiles of the first three ions have been scaled to
match that of \protect\zn2.  The scale factors are noted in the plots.
The \protect\s3\ profile has been shifted by $+2.2$ \protect\kms\ (see
text).  The \protect\s3\ profile is not plotted for $v_\odot > 5$
\protect\kms\ because of strong contamination with intermediate
negative velocity \protect\ion{Si}{2} \protect\wave{1190.416}
absorption components.  \label{fig:rholeo}}
\end{figure}

Toward \rholeo\ we also see that the \al3\ and \zn2\ profiles are
quite similar, showing two principal absorbing components (blends)
centered at velocities $v_\odot \approx -7$ and $+20$ \kms.  To show
this correspondence more clearly we plot the apparent column density
profiles (Savage \& Sembach 1991) of \ion{Zn}{2}, \s3, and \al3\ for
the sightline towards \rholeo\ in Figure \ref{fig:rholeo}.  Also shown
is the profile of the $^2P_{3/2}$ \ion{C}{2} fine structure level as
observed by the \ghrs\ G160M grating with a resolution of $\sim17$
\kms\ (FWHM).  The absolute velocity scale of the \c2star\
observations was determined through the use of a SPYBAL observations
as discussed in \S \ref{sec:data}.  The nominal wavelengths were
shifted by $-0.039$ \AA\ (or $-8.7$ \kms\ at \wave{1335.77}) based
upon the analysis of the SPYBAL observations.

The apparent column density profiles of \al3\ and \s3\ have shapes
very similar to the \zn2\ profile, and the alignment of the \al3\ and
\zn2\ component structure is excellent.  The \s3\ profile is
systematically shifted by $\sim-2.2$ \kms\ with respect to \zn2.  This
may be due to uncertainties in the absolute velocity scale; however,
Fitzpatrick \& Spitzer (1994) have found a similar offset ($-2.4$
\kms) in their \s3\ observations towards $\gamma^2$ Vel and suggested
a possible error in the rest wavelength of the transition.  Given the
similarities of the two profiles, we have applied a $+2.2$ \kms\ shift
to the observed \s3\ velocities in producing the profile in Figure
\ref{fig:rholeo}.  The profile for \c2star, which is likely tracing
thermal electrons that are responsible for exciting \ion{C}{2} to the
fine structure level (see SF93), also seems to trace the \zn2\
component structure very well, though the difference in resolution
between the two data sets make a direct comparison somewhat uncertain.

The detailed velocity correspondence between tracers of ionized and
primarily neutral gas seen towards the two high-$z$ stars in our
sample suggests the ionized gas is associated physically with the
neutral material.  {\em There is no kinematic evidence that the
ionized and neutral phases are spatially separated towards \rholeo.}
The good velocity correspondence between neutral and ionized gas
tracers could arise if \s3\ and \al3\ absorption were tracing ionized
edges of neutral clouds (e.g., McKee \& Ostriker 1977) or if the
clouds seen towards these high-latitude stars were partially ionized
with neutral and ionized tracers mixed (e.g., SF93).  The kinematic
profiles in our data do not allow us to distinguish between these two
scenarios, or other more complex arrangements of the two phases.  The
ionized and neutral media along the \hd18100\ sightline seem to show a
similarly close relationship, though we will not make as strong a
claim in this case given the lower resolution of the data.  The
ionization produced by decaying neutrinos as proposed by Sciama (1995,
1997) would result in partially ionized gas.  However, the decay
photons from neutrinos with $E_\gamma = 13.7\pm0.1$ eV (Sciama 1995)
are incapable of ionizing Al$^{+}$ and S$^{+}$.  Therefore other
sources of ionization (e.g., star light, X-ray background photons, or
collisional ionization) are required to produce the Al$^{+2}$ and
S$^{+2}$ we observe.

The close association of neutral and ionized tracers was also observed
along the sightline to \93521\ by SF93, a high-latitude star at
$z\approx1500$ pc from the Galactic plane.  The component structure
along the sightline towards \93521\ is more complex than that towards
\rholeo, but the correspondence between \c2star\ and tracers of
neutral gas (e.g., \ion{S}{2}) is very good.  Spitzer \& Fitzpatrick
interpreted the data for the \93521\ sightline to imply the free
electrons and neutral gas were cospatial and well-mixed.  Thus, they
argue for the existence of a partially ionized phase of the ISM at
high-$z$.

Recent \wham\ observations of the Perseus Arm have shown the intensity
ratio of [\ion{O}{1}] $\lambda$6300 to \Ha\ is in the range 0.01 to
0.04 (Reynolds \etal\ 1998a).  The weakness of [\ion{O}{1}] emission
implies the fractional ionization in the WIM of the Galaxy is very
high since the ionization of O and H are strongly coupled through
charge-exchange reactions.  Reynolds \etal\ comment that the observed
ratio implies that most of the observed emission from the Galactic WIM
cannot arise from partially-ionized gas along these sightlines, which
probe gas at distances from the plane $|z| \lesssim 300$ pc.  Similar
results have been obtained for the WIM in the edge-on galaxy NGC 891
(Dettmar \& Schulz 1992; Rand 1998).  The assumption by SF93 of a
partially-ionized phase of the ISM would seem to be contradicted by
these [\ion{O}{1}] measurements, although the \wham\ observations were
taken along a different path through the ISM and probe distances $z
\lesssim 300$ pc.

Further high-resolution observations of neutral and ionized gas
tracers will help to disentangle the connections between these phases
of the ISM.  Given the presence of \ion{C}{4} and \ion{N}{5}
absorption at similar velocities towards \hd18100, it is also possible
that these high-latitude sightlines are not tracing gas photoionized
by stellar radiation, but rather photoionization by cooling hot gas or
more complicated interactions between various phases of the ISM, such
as conductive interfaces or turbulent mixing layers (see SS96 and
references therein).  The contribution to the column densities of \s3\
and \al3\ by hot, collisionally ionized gas could compromise our
results and is discussed in \S \ref{subsec:collisional}.

Although the source of ionization of the WIM is not well understood,
in \S \ref{subsec:icfwim} we will derive the gas-phase abundances in
the ionized gas towards the high-latitude stars \rholeo\ and \hd18100\
by assuming the clouds to be photoionized by radiation from OB stars.

\section{GAS-PHASE ABUNDANCES IN IONIZED GAS}
\label{sec:abundances}

Deriving the gas-phase abundance of Al relative to S from the column
densities of \al3\ and \s3\ requires the application of an ionization
correction factor (ICF) to account for the unobserved ionization
stages of Al and S (primarily Al$^+$ and S$^+$).  The logarithmic gas
phase abundance normalized to solar, [Al/S]$_i$, where the subscript
$i$ denotes this value in the ionized gas, is related to the measured
column densities of \s3\ and \al3\ by
\begin{eqnarray}
 & [{\rm Al}/{\rm S}]_i \equiv 
	\log \{N({\rm Al}^{+2})/N({\rm S}^{+2}) \} 
	- \log \{ {\rm	Al}/{\rm S}\}_\odot  \nonumber \\
	&\  - \log \{ x({\rm Al}^{+2})/x({\rm S}^{+2}) \},
\label{eqn:ioncorrect}
\end{eqnarray}
where $x({\rm Al}^{+2}) \equiv N({\rm Al}^{+2})/N({\rm Al})$ and
$x({\rm S}^{+2}) \equiv N({\rm S}^{+2})/N({\rm S})$ are the ionization
fractions of Al$^{+2}$ and S$^{+2}$ in the ionized gas.  The ratio $\{
x({\rm Al}^{+2})/x({\rm S}^{+2}) \}^{-1}$ is the ICF, which we will
write ICF(Al$^{+2}$).  More generally in this work, ICF$(X^i) \equiv
\, \{ x(X^i)/x({\rm S}^{+2})\} ^{-1}$.

In this section we derive $x({\rm Al}^{+2})/x({\rm S}^{+2})$ for
moderate to low-density ionized gas near late-O/early-B stars.  We
then apply the ICFs to our measurements to derive gas-phase abundances
[Al/S]$_i$.

\subsection{CLOUDY Photoionization Equilibrium Models}
\label{subsec:cloudy}

We use the photoionization equilibrium code \cloudy\ (v90.04; Ferland
1996 and Ferland \etal\ 1998) to model the ionization and temperature
structure of diffuse, low-excitation \HII\ regions.  Our models assume
spherically symmetric nebulae excited by a single star.  We use a
volume filling factor $f$, which is the fraction of the volume filled
with constant hydrogen particle density $n_{\rm H}$ (in \percm3); the
rest of the space is assumed to be filled with very tenuous material
and in these models is treated as a vacuum.  From the
radially-averaged ionization structure of the nebula, we can derive
the ICF appropriate for direct measures of the column densities toward
the central ionizing stars.

We use ATLAS line-blanketed, LTE stellar atmosphere models (Kurucz
1991) as input spectra to the \cloudy\ models.  Our models follow the
temperature and ionization structure of the model \HII\ region from
0.3 pc distance from the exciting star to the point where the electron
density falls to $<5\%$ the ambient density $n_{\rm H}$ (i.e., $x({\rm
H^o}) = 0.95$).  While the models we present here assume solar
abundances with no dust opacity, we have found the inclusion of
sub-solar abundances of the refractory elements and the addition of
dust grain opacity does not significantly alter our conclusions.  This
is because highly refractory elements, i.e., Fe, Ni, Al, Cr, etc., are
not dominant nebular coolants.  Further, grain opacity tends to mimic
the absorbing characteristics of H; including this opacity therefore
removes the same ionizing photons as H (see Mathis 1986b).  While the
inclusion of photoelectric heating by dust and the absence of minor
nebular coolants may have a more pronounced effect on the predicted
emission line strengths, these thermal differences cause little or no
change in the predicted ionization fractions, and hence column
densities, of the species we are considering.  Detailed descriptions
of our use of \cloudy\ models to gain information regarding the
high-ionization species towards \mucol\ and the contribution of \HII\
region material to the neutral tracers towards this star can be found
in Brandt \etal\ (1998) and Howk \etal\ (1998), respectively.

We have computed a grid of model \HII\ regions with varying input
spectra and ambient densities.  Table \ref{table:cloudy_density} gives
the radially-averaged ionization fractions for several UV tracers of
photoionized gas in model \HII\ regions.  These models used ATLAS
model atmospheres with $T_{eff} = 33,000$ K, $\log L_* / L_\odot =
4.4$ and solar system abundances.  This stellar effective temperature
and luminosity are appropriate for \mucol\ or \zoph.  We have adopted
$f = 1.0$ in these models while varying the ambient density of the
gas.  Many authors define an ionization parameter, which is related to
the ratio of hydrogen-ionizing photon density, $n_\gamma$, to particle
density, $n_{\rm H}$, in the ionized region.  A traditional definition
of the ionization parameter (e.g., Mathis 1986b; Shields \& Kennicutt
1995) is
\begin{equation}
U \equiv \frac{L}{4 \pi  R_S^2 n_{\rm H} c},
\end{equation}
where $L$ is the hydrogen ionizing photon luminosity in photons
s$^{-1}$, and the Str\"{o}mgren radius is $R_S = [ 3L/(4\pi n_{\rm
H}^2 f \alpha_B) ]^{1/3}$, where $\alpha_B$ is the case B
recombination coefficient of H (Osterbrock 1989), and $f$ is again the
volume filling factor.  Using this definition $U \propto n_{\rm
H}^{1/3}$, and we find $3U = \langle n_\gamma / n_{\rm H} \rangle$,
where the average is over volume.  In this work we will primarily
adopt the equivalent definition given by Domg\"{o}rgen \& Mathis
(1994; hereafter DM94).  We write the alternate ionization parameter
$q$ as
\begin{equation}
q \equiv n_{\rm H} f^2 L_{50},
\label{eqn:ionparam}
\end{equation}
where $L_{50}$ is the stellar ionizing luminosity in units of
$10^{50}$ photons s$^{-1}$.  This definition is related to the
traditional $U$ by $q = 10^{-50} (36 \pi c^3/\alpha_B^2) U^3$, though
it removes the dependence on temperature that is hidden in $\alpha_B$
(Osterbrock 1989).  Bright high-density \HII\ regions are typically
described by models with $\log (q) \gtrsim -1.0$, while Domg\"{o}rgen
\& Mathis use $-4.0 \lesssim \log (q) \lesssim -3.0$ in modelling the
Galactic DIG.  We give the value of $\log (q)$ for each model in Table
\ref{table:cloudy_density} for comparison with the models of DM94,
Mathis (1986a) and others.  We also give the equivalent values of
$\log (U)$.  The values $U$ quoted assume $\alpha_B =
2.59\times10^{-13}$ cm$^3$~s$^{-1}$ appropriate for $T_e = 10^4$ K
(Osterbrock 1989).

In general the ionization and thermal structure of low-density models
is determined by the ionization parameter, the adopted gas-phase
abundances, and the shape of the ionizing continuum.  Table
\ref{table:cloudy_density} shows the ionization fraction of S$^{+2}$
is a function of the ionization parameter $q$; however the ratios of
the ionization fractions $x({\rm Al}^{+2})/x({\rm S}^{+2})$, $x({\rm
Si}^{+2})/x({\rm S}^{+2})$, and $x({\rm P}^{+2})/x({\rm S}^{+2})$ are
quite insensitive to changes in $q$ for a given stellar effective
temperature.  Further, these ratios, which give us the ICF for
determining the relative gas-phase abundances of Al, Si, and P to S,
are not large.  The greatest correction, for Al, is only of order
$\sim 0.2$ dex.  The ICF relating $N(\mbox{\fe3})/N(\mbox{\s3})$ to
[Fe/S]$_i$ is also relatively well-behaved with respect to the
ionization parameter, but we shall see that it shows substantial
variation with the input stellar effective temperature.

Table \ref{table:cloudy_temp} gives the ratios of ionization fractions
for $\log(q) = -4.0$ as a function of stellar effective temperature in
the range $27,000 \leq T_{eff} \leq 39,000$.  One can see that the
relative ionization fractions $x(X^i)/x({\rm S}^{+2})$ are more
sensitive to changes in the shape of the underlying continuum than to
changes in the ionization parameter $q$.  Even so, the spread of
values $x({\rm Al}^{+2})/x({\rm S}^{+2})$, $x({\rm Si}^{+2})/x({\rm
S}^{+2})$, and $x({\rm P}^{+2})/x({\rm S}^{+2})$ are still relatively
small.  Table \ref{table:cloudy_temp} also gives the average
ionization corrections appropriate for relating the ratio of the
various ionized species to their gas-phase abundances in the ionized
gas.  Two values are given: one representing the radially-averaged
physical properties of the nebula (ICF$_{Rad}$) and one the
volume-averaged properties (ICF$_{Vol}$).  The radially-averaged
values are appropriate for the sightlines that primarily probe
absorption through \HII\ regions towards the exciting stars.  The
volume-averaged ICFs are more appropriate for random interceptions of
unrelated \HII\ regions or the WIM.  The volume-averaged ICFs tend to
weight the outer portions of the model nebulae more strongly, and
therefore the volume-averaged values tend to favor lower stages of
ionization.  This can be seen in the difference between $ \log \langle
{\rm ICF(Si^{+3})}_{Vol} \rangle $ and $\log \langle {\rm
ICF(Si^{+3})}_{Rad} \rangle$ in Table \ref{table:cloudy_temp}: the
radially-averaged ICF suggests a greater fraction of Si is in
Si$^{+3}$ than in the volume-averaged case.  The ICFs given at the
bottom of Table \ref{table:cloudy_temp} are averages over the seven
stellar effective temperatures considered in our models with $\log (q)
= -4.0$ and $-2.0$.  Also given are the standard deviations about the
means.  The standard deviations give us a measure of how sensitive our
derived ICFs are to uncertainties in the effective temperature of the
ionizing radiation source, and we will use these values as estimates
of the errors in our derived ICFs.

\begin{figure}
\plotone{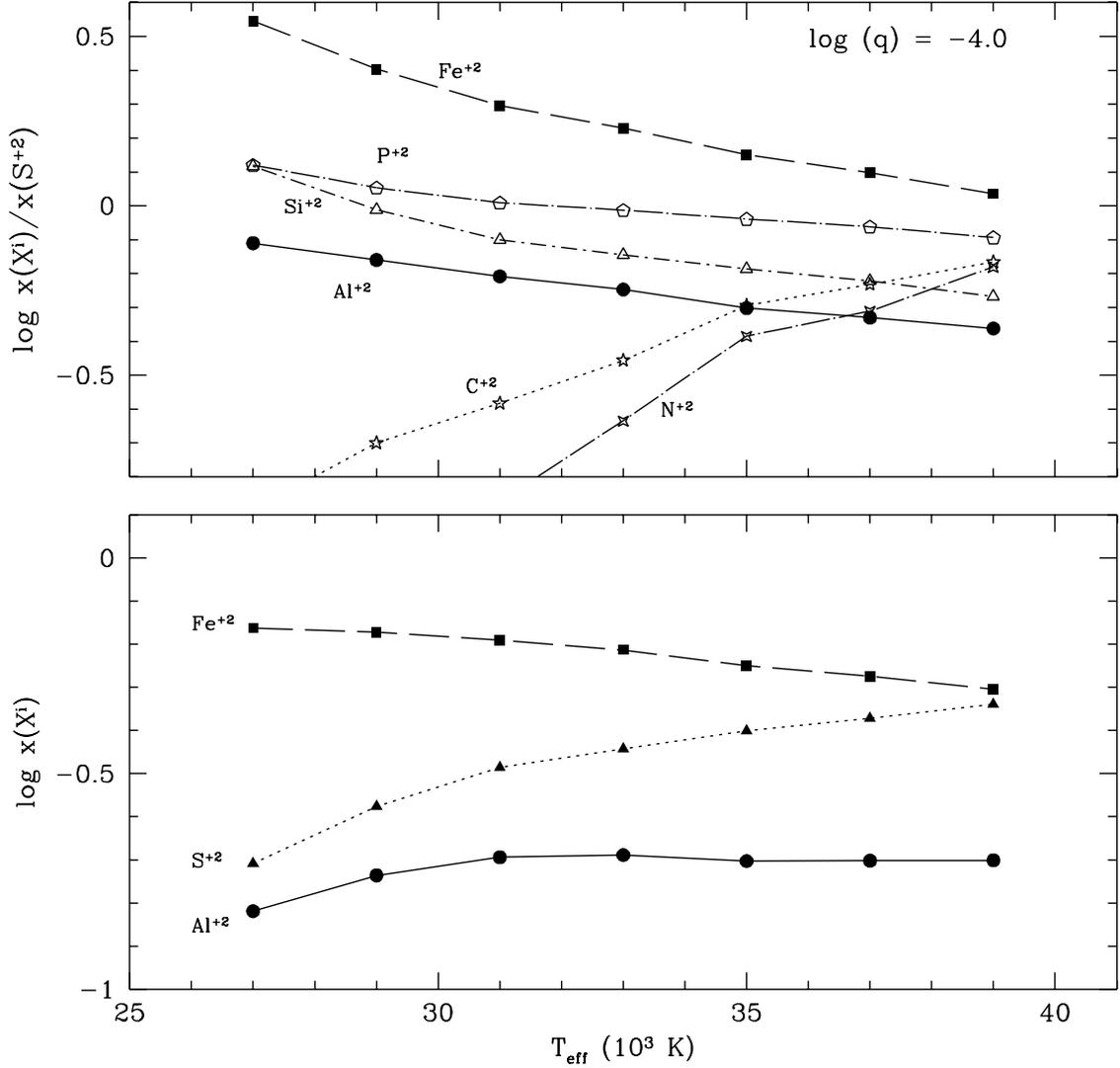}
\figcaption{The values of $\log x(X^i)/x({\rm S}^{+2}) = -\log {\rm
ICF}(X^i)$ for several tracers of ionized gas as a function of the
effective temperature, $T_{eff}$, of the ionizing star.  The
fractional ionization $x(X^i) \equiv N(X^i)/N(X)$ is derived using
column densities integrated from the star to the edge of the model
\HII\ region.  These data are for models characterized by the
ionization parameter $\log (q) = -4.0$.  Also shown in the bottom
panel is the behavior of the ionization fractions $x({\rm S}^{+2})$,
$x({\rm Fe}^{+2})$, and $x({\rm Al}^{+2})$ as a function of the
assumed stellar effective temperature.
\label{fig:cloudy_temp}}
\end{figure}

In Figure \ref{fig:cloudy_temp} we show $\log [x(X^i)/x({\rm
S}^{+2})]$ as a function of $T_{eff}$ for the ionized species $X^i =
{\rm Al}^{+2}, \ {\rm Fe}^{+2}, \ {\rm Si}^{+2},$ ${\rm P}^{+2}$,
${\rm C}^{+2}$, and ${\rm N}^{+2}$.  Also shown in the bottom panel of
this figure are the logarithms of $x({\rm S}^{+2}), \ x({\rm
Al}^{+2})$, and $x({\rm Fe}^{+2})$. It is clear that the ICFs relating
these ions to the gas-phase abundances of these elements (relative to
S) in the ionized gas are functions of the assumed stellar effective
temperatures.  However, for \al3, \si3, and \p3, the dependence on
$T_{eff}$ is relatively small, having a {\em total} spread of
$\lesssim 0.25$ dex over the range $29,000 \leq T_{eff} \leq 39,000$
K.  The bottom panel shows that even when S$^{+2}$ and Al$^{+2}$ are
not the dominant ionization stages of Al and S, the fractional
abundances of the two ions follow each other very well.  The $1\sigma$
uncertainties quoted for the ratios of ionic column densities using
absorption line spectroscopy are often $\sim0.1$ dex (e.g., Savage,
Cardelli, \& Sofia 1992; SF93; Howk \etal\ 1998), which is similar to
or greater than the standard deviations found for the predicted ICFs
given in Table \ref{table:cloudy_temp} for \al3, \si3, and \p3\
relative to \s3.

The ICF for \fe3\ (also \ion{C}{3} and \ion{N}{3}) shows a greater
dependence on the effective temperature of the underlying stellar
atmosphere than those for \al3, \si3, and \p3.  However, even
uncertainties on the order 0.2 to 0.3 dex for the value of
ICF(Fe$^{+2}$) can distinguish between depleted and non-depleted
abundance ratios for an element that is typically found to be as
heavily incorporated into grains as Fe.

A concern when interpreting these results is the accuracy of the
atomic data used in deriving these models.  In particular the atomic
data for Fe are a concern (e.g., Pradhan \& Bautista 1998).  The
atomic data adopted by \cloudy\ are discussed in Ferland (1996) and
Ferland \etal\ (1998).  In general the photoionization cross-sections
[using fits to the Opacity Project results by Verner \etal\ (1996)]
and radiative recombination rates (see references in Ferland \etal\
1998) are relatively secure.  The dielectronic recombination
coefficients $\alpha_{di}$ for recombinations into the first two ions
of Al are also relatively secure (Nussbaumer \& Storey 1986).  Though
$\alpha_{di}({\rm Al}^{+2})$ and $\alpha_{di}({\rm Al}^{+3})$ are
estimated, Ferland (1996) predicts the ionization balance of Al$^{\rm
o} - {\rm Al}^{+2}$ is relatively reliable.  The low-temperature
dielectronic recombination coefficients for the various ionic stages
of S, Si, and Fe are not well constrained (Ferland 1996; Ferland
\etal\ 1998).

\cloudy\ estimates the unknown low-temperature dielectronic
recombination coefficients of the first four ionization stages of
elements in the third and fourth rows of the periodic table by
adopting a mean of the rate coefficients for the first four ionization
stages of C, N, and O (Ali \etal\ 1991).\footnote{The adopted
coefficients for the first four ionization stages of third and fourth
row elements are: $\alpha_{di}(X^{\rm o}) = 3\times10^{-13}$;
$\alpha_{di}(X^{+}) = 3\times10^{-12}$; $\alpha_{di}(X^{+2}) =
1.5\times10^{-11}$; and $\alpha_{di}(X^{+3}) = 2.5\times10^{-11}$
cm$^{3}$ s$^{-1}$.  These are adopted from Nussbaumer \& Storey
(1983).}  We have disabled \cloudy's approximation of $\alpha_{di}$
for these elements to test its effects on the observed ratios.  The
ICFs derived for ${\rm Al}^{+2}$, ${\rm Si}^{+2}$, and ${\rm P}^{+2}$
differ in models with and without the assumptions regarding
dielectronic recombination by less than the standard deviations given
in Table \ref{table:cloudy_temp}.  For \fe3\ the models with
$\alpha_{di}({\rm Fe}^i) = 0.0$ give values close to that given in
Table \ref{table:cloudy_temp}.  However, the density dependence of
$x({\rm Fe}^{+2}) / x({\rm S}^{+2})$ is greatly increased.  The values
of ICF$({\rm Fe}^{+2})$ is less certain than the ICFs for the other
species considered here given their strong dependence on the shape of
the ionizing spectrum.  The same can be said for ICF$({\rm Si}^{+3})$.
The adopted error of $\sigma = 0.2$ to 0.3 dex for ICF(${\rm
Fe}^{+2}$) likely encompasses enough of parameter space to make our
error estimates not unreasonable.  We will make estimates for
[Fe/S]$_i$ where the data for \fe3\ are available, but it is important
to recognize the strong dependence of the ionization correction on
$T_{eff}$ and the possibly inappropriate atomic parameters for Fe.

\subsection{[Al/S]$_i$ and [Fe/S]$_i$ in H$\,${\small II} Regions}
\label{subsec:icfhii}

In this section we derive [Al/S]$_i$ and [Fe/S]$_i$ for the sightlines
probing \HII\ regions in the Galactic disk.  The results of \S
\ref{subsec:cloudy} give ICFs that are directly applicable to \HII\
regions about the stars \zoph, \xiper, \betasco, and \mucol.  The
distances to these stars are small enough that it may be reasonable to
assume their \HII\ regions dominate the column density of ions
considered here.  For \zoph\ this comparison is complicated by the
velocity structure, which suggests there are a number of processes at
work along this sightline.  A comparison of the integrated column
densities of \al3\ and \s3\ towards this star will not necessarily
provide us a good measure of the gas phase abundance [Al/S]$_i$.  As
discussed in \S \ref{subsec:hiiregions}, the sightline towards \mucol\
may be probing \HII\ region or WIM gas, or a mixture of both.  We will
assume here that the ionized gas along this sightline is probing an
\HII\ region.

Table \ref{table:ratios} presents the observed ratios of the
integrated column densities of \al3\ and \fe3\ (where available) to
\s3\ for our program stars.  Also given in this table are the derived
logarithmic abundances [Al/S]$_i$ and [Fe/S]$_i$.  The abundances
[Al/S]$_i$ and [Fe/S]$_i$ for the sightlines to \mucol, \xiper,
\betasco, and \zoph\ were derived using the ICFs appropriate for the
stellar effective temperatures of these stars, with the error
estimates as discussed in \S \ref{subsec:cloudy}.  We are assuming in
this approach that the measured material is mostly photoionized \HII\
region gas.  Given the velocity structure discussed in \S
\ref{subsec:zoph}, [Al/S]$_i$ in the \zoph\ \HII\ region may be as
high as $-1.0$.  The values [Al/S]$_i$ and [Fe/S]$_i$ presented in
Table \ref{table:ratios} for these disk sightlines suggest grains are
present in the \HII\ regions surrounding all of these stars.

Federman \etal\ (1993) present a \ghrs\ measurement of \p3\
\wave{1334.813} along the sightline to \zoph.  They find $\log
N(\mbox{\p3}) = 12.94 \pm 0.04$.  From the results presented in Table
\ref{table:cloudy_temp}, we see $\log {\rm ICF}({\rm P}^{+2}) = +0.01
\pm 0.06$.  We thus find [P/S]$_i = -0.11 \pm 0.07$ for the ionized
gas towards \zoph.  This value assumes the integrated sightline column
densities of P and S are tracing the same regions, which may be
incorrect given the velocity structure towards this star (\S
\ref{subsec:zoph}).  The element P is thought to be very lightly
depleted.  In their study of the neutral gas abundances along the
\zoph\ sightline, Savage \etal\ (1992) find [P/H]$\, = -0.23$ for the
warm neutral cloud centered at $v_\odot = -27$ \kms\ (their component
A).  For the sightline towards \mucol, Howk \etal\ (1998) derive
[P/S]$\, = -0.03 \pm 0.03$ in the low-velocity absorbing complex
centered at $v_\odot = +23$ \kms, which shows abundance patterns
similar to component A towards \zoph.  The [P/S]$_i$ derived here for
the \zoph\ sightline is consistent with previous [P/S] measurements
for neutral material: P is lightly depleted if at all.

\subsection{[Al/S]$_i$ in the WIM}
\label{subsec:icfwim}

In this subsection we consider the gas-phase abundances [Al/S]$_i$ in
the halo WIM.  The interpretation of the
$N(\mbox{\al3})/N(\mbox{\s3})$ measurements along the sightlines to
\rholeo\ and \hd18100\ is subject to uncertainties regarding the
ionization of the WIM and contamination from collisionally ionized gas
(see \S \ref{subsec:collisional}).  The long path-lengths through the
WIM and the likelihood that any \HII\ regions around these stars are
in very low-density environments imply that a large fraction of the
observed \al3\ and \s3\ along these sightlines arises in the WIM of
our Galaxy.  Rho Leo should lie above $\sim 60\%$ of the WIM of the
Galaxy, while \hd18100\ should lie above almost all of the high-$z$
layer of ionized material (Reynolds 1989; Savage \etal\ 1990).  Though
the ionization source of the diffuse ionized material is not well
known, the power requirements ionization by OB stars may be a viable
mechanism (Reynolds 1984; DM94).  Here we estimate [Al/S]$_i$ towards
\rholeo\ and \hd18100\ {\em for the case where the WIM of the Galaxy
is photoionized by OB stars.}

We will follow DM94 and model the WIM of the Milky Way using
low-density \HII\ region photoionization models.  The ratio $x({\rm
Al}^{+2})/x({\rm S}^{+2})$ derived for our \HII\ region models is
relatively insensitive to the ionization parameter and is not heavily
dependent on the shape of the input ionizing spectrum; we therefore
believe it is appropriate to apply the results derived in \S
\ref{subsec:cloudy} to the WIM of the Galaxy.  Our observations of
ionized gas towards \rholeo\ and \hd18100\ probe the WIM and thus are
not a simple radial integration through an \HII\ region.  A
volume-averaged set of ICFs are more appropriate for application to
the WIM than the radially-averaged values used in the previous
subsection (DM94).  Volume-averaged ICFs from the models treated in \S
\ref{subsec:cloudy} are given in the last row of Table
\ref{table:cloudy_temp}.  We see that there are significant
differences in the ICFs derived for most of the ions studied here when
adopting volume- versus radially-averaged values.

We do not know well the shape of the ionizing spectrum appropriate for
the WIM. We will continue by adopting the mean of the volume-averaged
values ICF$({\rm Al}^{+2})$ for our models.  These values are given in
Table \ref{table:cloudy_temp} as ICF$_{Vol}$.  The values [Al/S]$_i$
given for \hd18100\ and \rholeo\ in Table \ref{table:ratios} are
therefore derived assuming $\log \{x({\rm Al}^{+2})/x({\rm S}^{+2})\}
= -0.37 \pm 0.07$, or $\log {\rm ICF}({\rm Al}^{+2}) = +0.37$ in the
WIM.  For the sightline towards \mucol, the value [Al/S]$_i$ given in
Table \ref{table:ratios} was derived assuming $\log {\rm ICF(Al}^{+2})
= +0.25\pm0.07$ in an \HII\ region about this star; if this sightline
instead probes the WIM along this direction, the gas-phase abundance
could be ${\rm [Al/S]}_i = -0.66\pm0.08$.

We can estimate values for $\log x({\rm S^{+2}})$ along our two
high-latitude sightlines.  Assuming $N_e = 7\times10^{19} / \sin |b|$
\percolumn\ (Reynolds 1991b), and accounting for the 40\% of this
value expected to reside beyond \rholeo, the predicted electron column
densities along the high-latitude sightlines are $\log N_e = 19.7$ and
$19.9$ for \rholeo\ and \hd18100, respectively.  Assuming $N_e \approx
N({\rm H}^+)$ and the solar abundance of S/H (Anders \& Grevesse
1989), for \rholeo\ and \hd18100\ we find $\log x({\rm S^{+2}}) \sim
-1.3$ and $-0.9$, respectively.  Compared with our {\em
volume-averaged} results for models having $\log (q) = -4.0$, this
implies the characteristic ``effective temperature'' for the ionizing
radiation field of $28,000 \lesssim T_{eff} \lesssim 33,000$ K is
appropriate for these sightlines.\footnote{It is important to note
that we are using the {\em volume-averaged} ICFs here, rather than the
radially-averaged values given in Table \ref{table:cloudy_temp}.  For
models having $\log (q) = -4.0$ we find $\log x({\rm S^{+2}}) = -1.38,
\ -1.16, \ -0.99, \ -0.84, -0.77$ and $-0.53$ for $T_{eff} = (27, 29,
31, 33, 35, 37, {\rm and} \ 39)\times 10^{3}$ K, respectively, in the
volume-averaged model results.}  This result is consistent with the
limits derived from observations of diffuse \ion{He}{1} recombination
radiation (Reynolds \& Tufte 1995).

The \al3\ and \s3\ data presented here, when compared with our model
results, are inconsistent with a solar relative abundance of Al to S
in the WIM.  We have assumed that the intrinsic or cosmic abundance
(gas$+$dust) of Al to S is given by the solar system values.  Even in
the presence of mildly sub-solar metallicities, $[X/{\rm H}] \gtrsim
-1.0$, the abundance of Al relative to S is expected to be quite close
to the solar abundance.  Though Al is an odd-Z element, it is
primarily produced in the C-,Ne-burning stages of massive stars.
Thus, S and Al are both deposited into the ISM by Type II supernovae.

In their study of elemental abundances in solar neighborhood low-mass
stars, Edvardsson \etal\ (1993) derive abundances for Al as well as
several $\alpha$-elements.  The behavior of [Al/Fe] versus [Fe/H] in
their dataset mimics that of the $\alpha$-elements.\footnote{The
element S is not included in their dataset, though see Wheeler, Sneden
\& Truran (1989) and references therein.}  For their sample of 189
stars, all having [Fe/H]$\, \gtrsim -1.0$, the average value of
[Al/$\alpha$] is $+0.02\pm0.07$, where $\alpha$ in this context
represents an average over the elements Mg, Si, Ca, and Ti.  The yield
of Al relative to the $\alpha$-elements does depend upon the initial
stellar abundances, but this effect is negligible for the range of
metallicities of the Edvardsson \etal\ sample (and for our halo
clouds).  We do not expect the intrinsic ratio of Al/S, as long as the
metallicity of the gas is $\gtrsim0.1$ solar, to be significantly
different than the solar-system value adopted here.  Therefore the
values of [Al/S]$_i < 0.0$ presented here for the WIM are not due to
nucleosynthetic effects and imply the existence of dust in this phase
of the ISM.

\subsection{Collisionally Ionized Gas and its Effects on [Al/S]$_i$}
\label{subsec:collisional}

If hot, collisionally ionized gas is present along the sightlines
studied here, its imprint on the ionization balance of the gas could
cause us to misinterpret the column density ratios given in Table
\ref{table:ratios}.  The collisional ionization equilibrium models of
Sutherland \& Dopita (1993), for example, suggest that for gas with
temperatures $T \sim 50,000$ K ($\log T = 4.7$), $\log[x({\rm
Al^{+2}}) / x({\rm S^{+2}})] = -0.93$.  Therefore, if one of our
sightlines were dominated by gas in collisional ionization equilibrium
at this temperature, with [Al/S]$_i = 0$, i.e., no depletion, we would
erroneously derive [Al/S]$_i \approx -0.5$ by assuming the value of
$\langle ICF \rangle _{Vol}$ given in Table \ref{table:cloudy_temp}.
While the assumption of collisional ionization equilibrium is not
likely to be valid, this calculation suggests the presence of gas with
$T \gtrsim 30,000$ K along our sightlines could compromise our
conclusions.  In this section we will give evidence suggesting, for
most of our sightlines, it is unlikely our results are strongly
affected by contamination from collisionally ionized gas.

As a first step for approaching the problem, we have collected column
density measurements of \ion{Si}{4}, and the ratio of \ion{Si}{4} and
\ion{S}{3} column densities, for all of our sightlines in Table
\ref{table:si4}.  Several of the \ion{Si}{4} column densities given in
Table \ref{table:si4} are taken from literature measurements using the
GHRS (with references given in the table); others were derived from
archival GHRS spectra using the techniques described in \S 2.  With an
ionization potential of 33.5 eV, the ion \ion{Si}{4} is potentially a
tracer of collisionally ionized gas with T in the range
$(0.3-1)\times10^5$ K, though \ion{Si}{4} can also be produced in warm
gas via photoionization by hot stars.

In the following discussion we consider the observational constraints
regarding collisional ionization for each sightline.  

{\em \zoph\ --} The absorption towards \zoph, as discussed in \S
\ref{subsec:zoph} and shown in Figure \ref{fig:zoph}, shows evidence
for changes in the ionization structure as a function of heliocentric
velocity.  The gas at $v_\odot \lesssim -15$ \kms\ is characterized by
large amounts of \ion{Si}{4} absorption and relatively low amounts of
\al3.  Higher velocity gas along this sightline, $-13 \lesssim v_\odot
\lesssim -5$ \kms\ shows relatively little \ion{Si}{4} and higher
ratios of \al3\ to \s3.  The peak of the \ion{Si}{4} profile may trace
gas that is collisionally ionized.  Our \cloudy\ models, e.g., Table
\ref{table:cloudy_density}, cannot explain the large value of the
ratio $ \log [ N(\mbox{\ion{Si}{4}})/N(\mbox{\ion{S}{3}}) ] \approx
-1.8$ observed near $v_\odot \approx -20$ \kms.  However, Sembach
\etal\ (1994) note that the \ion{Si}{4} line width implies gas with
$T<5\times10^4$ K.  This temperature limit and the absence of
associated \ion{C}{4} absorption lead Sembach \etal\ to suggest the
\ion{Si}{4} absorption is due to photoionized gas in an expanding
\HII\ region.  The gas between $v_\odot = -8$ and $-13$ \kms\ is also
likely photoionized gas in the \HII\ region about \zoph.  This gas
shows values of $ \log [ N(\mbox{\ion{Si}{4}})/N(\mbox{\ion{S}{3}}) ]
\lesssim -2.0$, which can be generally explained by photoionization
models.  The profile width of the \al3\ absorption measured by Sembach
\etal\ (1994) implies gas with $T<3\times10^4$ K.  Morton (1975) also
finds absorption from \ion{N}{2}, \ion{N}{2}$^*$, \ion{N}{2}$^{**}$ at
$v_\odot = -8$ \kms, suggesting the presence of (photo)ionized gas and
a significant density of electrons at these velocities.  Our earlier
suggestion that a value of $ \log [
N(\mbox{\ion{Al}{3}})/N(\mbox{\ion{S}{3}}) ] \approx -2.0$ is most
appropriate for the photoionized \HII\ region gas towards \zoph\ is
based upon the observed ratio at velocities corresponding to the peak
of the \al3\ (and \ion{N}{2}) absorption.  Reynolds (1988b) has
observed \Ha\ emission centered at $v_\odot = -13 \pm 1$ \kms\ along
this sightline, which is quite near the peak of the \s3\ absorption.
The temperature implied by an analysis of the breadth of the
[\ion{S}{2}] and \Ha\ emission profiles is $T = 6700\pm700$ K
(Reynolds 1988b).  Thus if we associate the \Ha\ and [\ion{S}{2}]
emission with the absorption seen in \s3, \al3, and \ion{N}{2}, the
temperatures are too low to produce significant amounts of these ions
through collisional ionization.

{\em \xiper\  --} The sightline towards \xiper\ shows a complex of \Ha
-emitting regions at several different velocities. Reynolds (1988b)
finds ionized regions excited by this star at velocities $v_\odot =
+3, \ +7, \ +12,$ and $+14$ \kms.  Figure \ref{fig:xiper} shows the
$N_a(v)$ profiles of
\s3, \al3, and \ion{Si}{4} towards \xiper\ and the ratios of the
latter two ions to \s3, as Figure \ref{fig:zoph} did for \zoph.  The
vertical dashed lines in this figure represent the velocities at which
Reynolds (1988b) detects \Ha\ emission, though shifted by $-2.2$ \kms.
The peaks in the $N_a(v)$ profile for \s3\ give a good match to the
components detected in emission by Reynolds.  In his analysis,
Reynolds (1988) used the emission from [\ion{S}{2}] and
\Ha\ to separate the thermal and non-thermal components of the
observed velocity widths.  He found that the emitting regions excited
by \xiper\ had temperatures between $T \approx 5,000$ and $11,000$ K,
consistent with expectations for photoionized gas.  We associate the
emitting regions studied by Reynolds with the material seen in
\s3\ and \al3\ absorption in our GHRS spectra.  At the implied
temperatures, the contribution to the \s3\ and \al3\ column densities
from collisionally ionized gas should be negligible.  The ICFs derived
using photoionization models should therefore be appropriate for this
material.

{\em \betasco\ --} The derived $\log
N(\mbox{\ion{Si}{4}})/N(\mbox{\ion{S}{3}})$ towards \betasco\ is quite
similar to the measured values along the \zoph\ and \xiper\
sightlines.  Unfortunately, we have virtually no information on the
velocity structure of the \s3\ absorption profile given the low
resolution of the G160M data.  The logarithmic ratio of $\log
N(\mbox{\ion{Si}{4}})/N(\mbox{\ion{Al}{3}})$ is in the range $-0.2$ to
0.0 over the velocity range of \al3\ absorption.  The profile-weighted
average velocities (Sembach \& Savage 1992) for \al3\ and \ion{Si}{4}
are $\langle v_\odot \rangle = -13.9\pm0.6$ and $-11.4\pm0.5$ \kms,
respectively.  It is tempting to use the similarities between the
\zoph\ and \betasco\ sightlines to argue that collisionally ionized
gas is not confusing our analysis of this sightline.  The neutral gas
absorption profiles towards these stars are very similar (e.g., the
\zn2\ profiles of Figure \ref{fig:spectra}) and the two stars lie at
approximately the same distance from the sun and from the Galactic
mid-plane.  However, the differences in spectral type would suggest
that in the case of purely photoionized gas and equal gas densities,
the ratio $N(\mbox{\ion{Si}{4}})/N(\mbox{\ion{S}{3}})$ should be
smaller for \betasco, i.e., for the star with lower $T_{eff}$.  Also,
we have seen (\S \ref{subsec:zoph}) that there are some peculiarities
along the \zoph\ sightline, possibly including the presence of
collisionally ionized gas at velocities $v_\odot \lesssim -15$ \kms.
We cannot rule out contributions from collisionally ionized gas to the
\s3\ and \al3\ absorption towards \betasco.

{\em \mucol\ --} For the sightline towards \mucol, measurements with
the \wham\ Fabry-Perot spectrometer (Reynolds \etal\ 1998) show the
\Ha\ emission is at velocities consistent with the \s3\ absorption
profile.  The \wham\ measurements, with a 1$^\circ$ beam, are well fit
by a gaussian with $\langle v_\odot \rangle = 22.9\pm0.5$ \kms\ and $b
= 17.0 \pm 1.5$ \kms\ (M. Haffner, private communication).  The
breadth of the profile suggests that the gas traced by this emission
has $T \lesssim 18,000$ K; in collisional ionization equilibrium
models of gas at this temperature \s3\ and \al3\ represent less than
1\% of the total S and Al abundances (Sutherland \& Dopita 1993).  The
profile-weighted average velocity (Sembach \& Savage 1992) of the \s3\
is $\langle v_\odot \rangle = 23.2\pm0.3$ \kms, with the \al3\
\twowave{1855 and 1862} lines having $\langle v_\odot \rangle =
21.4\pm1.0$ and $23.9\pm0.9$ \kms, respectively.  Due to the
similarity of the velocities, we associate the \Ha -emitting gas with
the \s3\ and \al3\ absorption.  The temperature limits set by the
\wham\ spectra of this sightline suggest that the contribution of hot,
collisionally ionized gas to the \s3\ and \al3\ column densities is
small.

{\em \rholeo\ --} The archival GHRS spectra of \ion{Si}{4} towards
\rholeo\ show no evidence for absorption in the velocity range covered
by the \s3\ absorption studied here (i.e., $-19 \leq v_\odot \leq +2$
\kms).  The $2\sigma$ upper limit to $N(\mbox{\ion{Si}{4}})$ given in
Table \ref{table:si4} implies $\log
[N(\mbox{\ion{Si}{4}})/N(\mbox{\ion{S}{3}})] \leq -2.1$ in this
velocity range, assuming $b_{\rm Si IV} \approx 10$ \kms.  Though
there is \ion{Si}{4} absorption overlapping the higher-velocity \al3\
absorption, it appears to be significantly broader than the \al3\
absorption.  We believe that the lack of detectable \ion{Si}{4}
absorption at the velocities of the stronger absorbing complex seen in
\al3\ (and \s3), with limits that place the ratio
$N(\mbox{\ion{Si}{4}})/N(\mbox{\ion{S}{3}})$ at levels well below
those detected along our \HII\ region sightlines, suggests
collisionally ionized gas is not an important contributor to the
column densities of \al3\ and \s3\ towards \rholeo.

{\em \hd18100\  --} The sightline towards the star \hd18100\ has been
studied by Sembach
\& Savage (1994), and their values of $N(\mbox{\ion{Si}{4}})$ are
given in Table \ref{table:si4}.  As mentioned in \S
\ref{subsec:hilatitude}, the velocities of \s3\ and \al3\ along this
sightline are quite similar to those of the low ions (e.g.,
\ion{Zn}{2} and \ion{Mn}{2}), as well as those of the highly-ionized
species \ion{Si}{4} and \ion{C}{4}.  Indeed, Figure 4 of Sembach \&
Savage (1994) shows that the profiles of \al3, \ion{Si}{4}, and
\ion{C}{4} are quite similar as viewed with the resolution of the GHRS
G160M grating.  Though the breadth of the profiles increases with
ionization potential along this sequence, the increase seems mostly to
occur towards negative velocities.  It is possible that the highly
ionized species discussed by Sembach \& Savage have more complex
velocity structures that would reveal themselves at higher resolution
[the 1.4 \kms\ resolution \ion{Ca}{2} profile presented by SS96 and
Ryans \etal\ (1996) shows evidence for many closely-spaced absorbing
components along this sightline].  However, we have no evidence with
which to rule out the possibility that the observed columns of
\s3\ and \al3\ are produced in collisionally ionized gas.  
The \ion{Si}{4} and \ion{C}{4} absorption profiles, which are at the
same velocities as the lower ionization species, suggest that such
contamination may indeed be a real problem for this sightline.  That
the ratio of $N(\mbox{\ion{Si}{4}})/N(\mbox{\ion{S}{3}})$ along this
sightline is the highest in our sample also suggests collisionally
ionized gas may be playing a role in determining the ionization
balance in the gas we are studying.  Thus our application of the ICFs
derived from photoionization for this sightline may be inappropriate.

In summary it appears, with the exception of the sightline to \hd18100
and possibly \betasco, collisional ionization does not likely explain
the origin of the \al3\ and \s3.  In the case of \hd18100, the
velocity overlap of the absorption due to low, moderate, and highly
ionized species, as well as the large column density of \ion{Si}{4}
relative to \s3, raises the possibility that much of the observed
\al3\ and \s3\ along this sightline could arise in collisionally
ionized gas.

\section{DISCUSSION}
\label{sec:discussion}

The \al3\ and \s3\ column densities, when combined with the results of
our photoionization modelling, suggest that Al-bearing dust is indeed
present in the ionized gas along the sightlines considered here.  For
\HII\ region gas this result is not surprising.  The existence of dust
in \HII\ regions has been implied by several independent lines of
reasoning (see Osterbrock 1989, Chap. 7; Mathis 1986b; Peimbert \&
Goldsmith 1972), though little can be discerned about the composition
of the grains or the degree to which dust grains are processed in the
ionized ISM using these methods.

Our measurements of [Al/S]$_i$ for the nearby sightlines that probe
the \HII\ regions surrounding the stars \zoph, \xiper, \betasco, and
\mucol\ give us a measure of the incorporation of Al into dust grains
in these nebulae.  Assuming [Al/S]$_i = {\rm [Al/H]}_i$, the
dust-phase abundances of Al in the ionized material along these
sightlines, in units of atoms per million hydrogen, are $10^6 ({\rm Al
/ H})_d \equiv 10^6({\rm Al / H})_\odot - 10^6({\rm Al / H})_i =
2.8\pm0.6, \ 2.6\pm0.5, \ 2.7\pm0.8,$ and $2.5\pm0.5$ for \zoph,
\xiper, \betasco, and \mucol, respectively.  These values are an order
of magnitude less than the dust-phase abundance of Fe, Si, or Mg in
the warm neutral ISM (Savage \& Sembach 1996b; Howk \etal\ 1998).  The
dust-phase abundances of Al presented here are similar to the values
derived for Ni in the WNM, which has a similar solar system abundance
(Savage \& Sembach 1996b; Howk \etal\ 1998).

Unfortunately, little is known about the gas-phase abundance of Al in
the WNM of the Galaxy due to the great strength of the only available
\ion{Al}{2} transition at \wave{1670}.  Barker \etal\ (1984) have
performed a curve-of-growth analysis of interstellar \ion{Al}{2}
absorption in the WNM using data from the \copernicus\ observatory.
However, for many of their sightlines the absorption due to Al is far
up the flat portion of the curve of growth, making the uncertainties
in the determination of the Al column densities quite large.  In
deriving the column density of \ion{Al}{2}, Barker \etal\ employ an
empirical curve of growth derived from \ion{Si}{2}.  However, since Al
and Si are expected to have different gas-phase abundances from
component to component due to changing depletion effects, it is likely
that \ion{Si}{2} and \ion{Al}{2} have somewhat different curves of
growth.  Jenkins (1983), with several important caveats summarized at
the beginning of his paper, has compared the equivalent widths of
\ion{Al}{2} \wave{1670} with the similarly strong \ion{Si}{2}
transition at \wave{1304}.  While the nature of his comparison tends
to heavily weight low-column density intermediate velocity features,
Jenkins finds no evidence for changing [Al/Si] with $z$-height of the
probe star with the implied value [Al/Si]$\, \approx -0.2$.  This is
near the upper range found by Barker \etal\ (1984), but is roughly
consistent with their values.  Again, there are many uncertainties
with Jenkins' approach, among them that the \ion{Al}{2} and
\ion{Si}{2} transitions are likely both on the flat part of the curve
of growth, which may hide significant changes in the relative
abundances.

For comparison with our results, we look to those sightlines with the
least saturated lines in the Barker \etal\ survey.  For these
sightlines Barker \etal\ consistently find [Al/H]$\, \approx -1.1$ to
$-1.0$.  This selection is strongly biased towards low-density
sightlines.  If we take [Al/S]$_i \approx -1.0$ as representative of
the \HII\ region sightlines in our sample, this suggests the
refractory grain material is not heavily affected by the conditions in
low-density \HII\ regions.  However, the validity of this comparison
is somewhat suspect given the uncertainties and biases in the Barker
\etal\ results.

Our estimates of [Fe/S]$_i$ in three of the sightlines considered here
(see Table \ref{table:ratios}) yield values roughly consistent with
those for [Fe/H]$_i$ in the Orion nebula from a variety of authors
(Osterbrock \etal\ 1992; Peimbert \etal\ 1993; Baldwin \etal\ 1996;
Rodr\'{\i}guez 1996; and Rubin \etal\ 1997) and suggest Fe is
incorporated into grains in these ionized regions.  The relative
values [Al/S]$_i$ versus [Fe/S]$_i$ suggest that the fraction of Al
incorporated into grains is consistent with that of Fe or a bit less.
For the sightlines towards \zoph\ and \mucol\ we can directly compare
the gas-phase abundances of Fe derived for the neutral and ionized
sightlines.  In each case, the velocity of the ionized gas is closest
to the components that show the lowest value of [Fe/H], which are
[Fe/H]$\, = -2.4$ for \zoph\ (Savage \etal\ 1992) and [Fe/H]$\,
\approx \,$[Fe/S]$\, = -1.3$ for \mucol\ (Sofia \etal\ 1993; Howk
\etal\ 1998).  Using the values of [Fe/S]$_i$ given in Table
\ref{table:ratios}, we find that $\sim3\%$ and $\sim9\%$ of the
dust-phase Fe in the neutral gas has been returned to the gas-phase in
the ionized gas (with a factor of two uncertainty).  Our measurements
of significantly sub-solar Al and Fe abundances in \HII\ regions
suggest the processing of grains in low-excitation \HII\ regions is
not much different than the destruction that occurs in the WNM of the
Galaxy.

The implications of dust within \HII\ regions have been discussed by
several authors (e.g., Sheilds \& Kennicutt 1995; Henry 1993; McGaugh
1991; Aannestad 1989; Mathis 1986b).  The most important effects are
caused by the incorporation of possibly important coolants into the
solid phase (Shields \& Kennicutt 1995), the change in the thermal
balance due to photoelectron emission (heating) and far-infrared
thermal dust emission (cooling), and the competition of the dust
opacity with H and He for ionizing photons (Mathis 1986b).  McGaugh
(1991) has shown that the existence of dust can significantly alter
the Balmer line strengths expected from an \HII\ region.  This effect
comes about because the dust is able to absorb a significant number of
photons that would otherwise go towards ionizing H (Mathis 1986b).  As
a consequence McGaugh suggests calculations of the star formation
rates using only Balmer line intensities may underestimate the number
of ionizing stars present.  Also, the true volume of a dusty \HII\
region will be smaller than the dust-free Str\"{o}mgren volume.  For
low-density \HII\ regions like those studied here, however, Mathis
(1986b) has shown that the effects of dust absorption may not provide
a significant optical depth to ionizing photons.  Mathis also points
out that although dust is an additional source of opacity, the opacity
due to dust does not affect the ionization balance of most species in
\HII\ regions given its similarity to the H opacity.  Shields \&
Kennicutt (1995) discuss the important effects of dust on emission
line strengths from metal rich ($Z>Z_\odot$) \HII\ regions,
particularly those found near the centers of galaxies.

\begin{figure}
\plotone{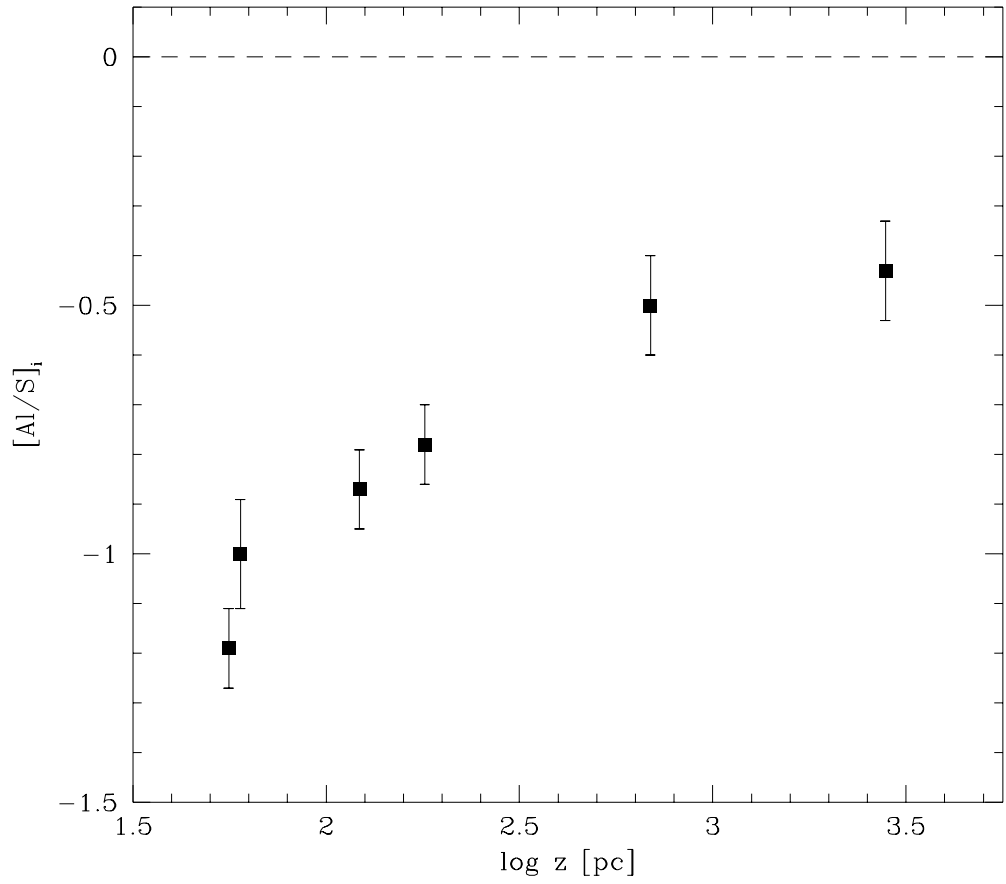} 
\figcaption{The behavior of [Al/S]$_i =\log ({\rm Al/S})_i - \log
({\rm Al/S})_\odot$, as a function of $z$-height of the probe star.
Since S is not depleted onto grains $[{\rm Al/S}]_i \approx [{\rm
Al/H}]_i$ gives a measure of the fraction of Al incorporated into dust
grains in the ionized gas.
\label{fig:deplz}}
\end{figure}

Sembach \& Savage (1996) have shown that in general, the gas-phase
abundances of elements increases as one moves from the disk to the
halo of our galaxy.  This suggests an increasing degree of
(incomplete) grain destruction with increasing height above the plane
of the Galaxy.  In Figure \ref{fig:deplz} we show the gas-phase
abundances [Al/S]$_i$ as a function of $z$-distance of the observed
stars.  Since S is generally not depleted [Al/S]$_i$ should closely
follow [Al/H]$_i$.  There is a general trend of increasing gas-phase
abundance of Al in the ionized gas with increasing height above the
plane of the Galaxy.  This is qualitatively consistent with the
behavior observed by Sembach \& Savage for the WNM.

\begin{figure}
\plotone{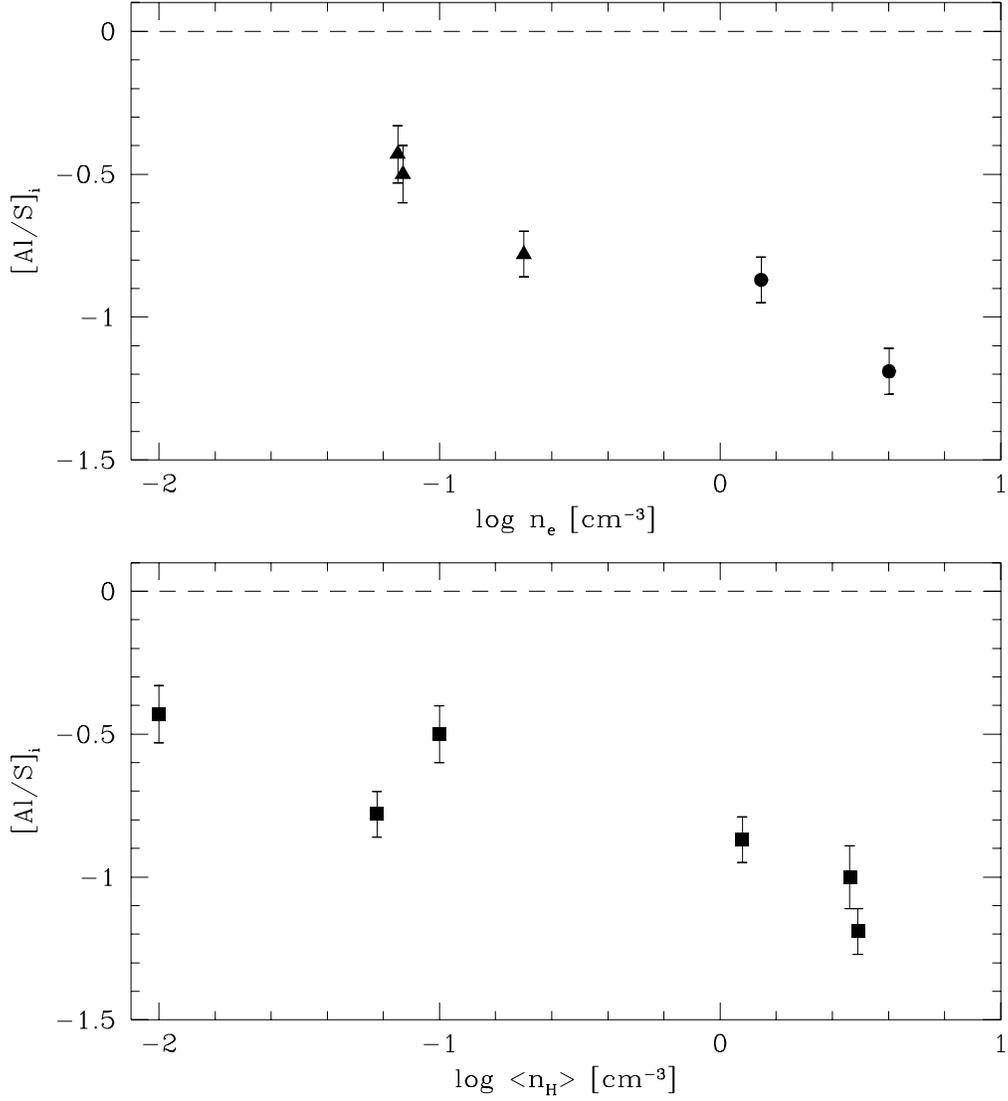} 
\figcaption{The behavior of [Al/S]$_i \equiv \log ({\rm Al/S})_i -
\log ({\rm Al/S})_\odot$ as a function of average density.  Since S is
not depleted onto grains $[{\rm Al/S}]_i \approx [{\rm Al/H}]_i$ gives
a measure of the fraction of Al incorporated into dust grains in the
ionized gas.  The bottom panel shows [Al/S]$_i$ versus the average
sightline neutral density $\langle n_{\rm H} \rangle \equiv
[N(\mbox{\HI}) + 2N({\rm H_2})]/d$.  The top panel shows [Al/S]$_i$
versus the electron density $n_e$.  The filled circles in the top
panel represent determinations of rms electron densities $\langle
n_e^2 \rangle^{1/2}$, while the triangles are for determinations of
average electron densities $\langle n_e
\rangle$. \label{fig:depldens}}
\end{figure}

\begin{figure}
\plotone{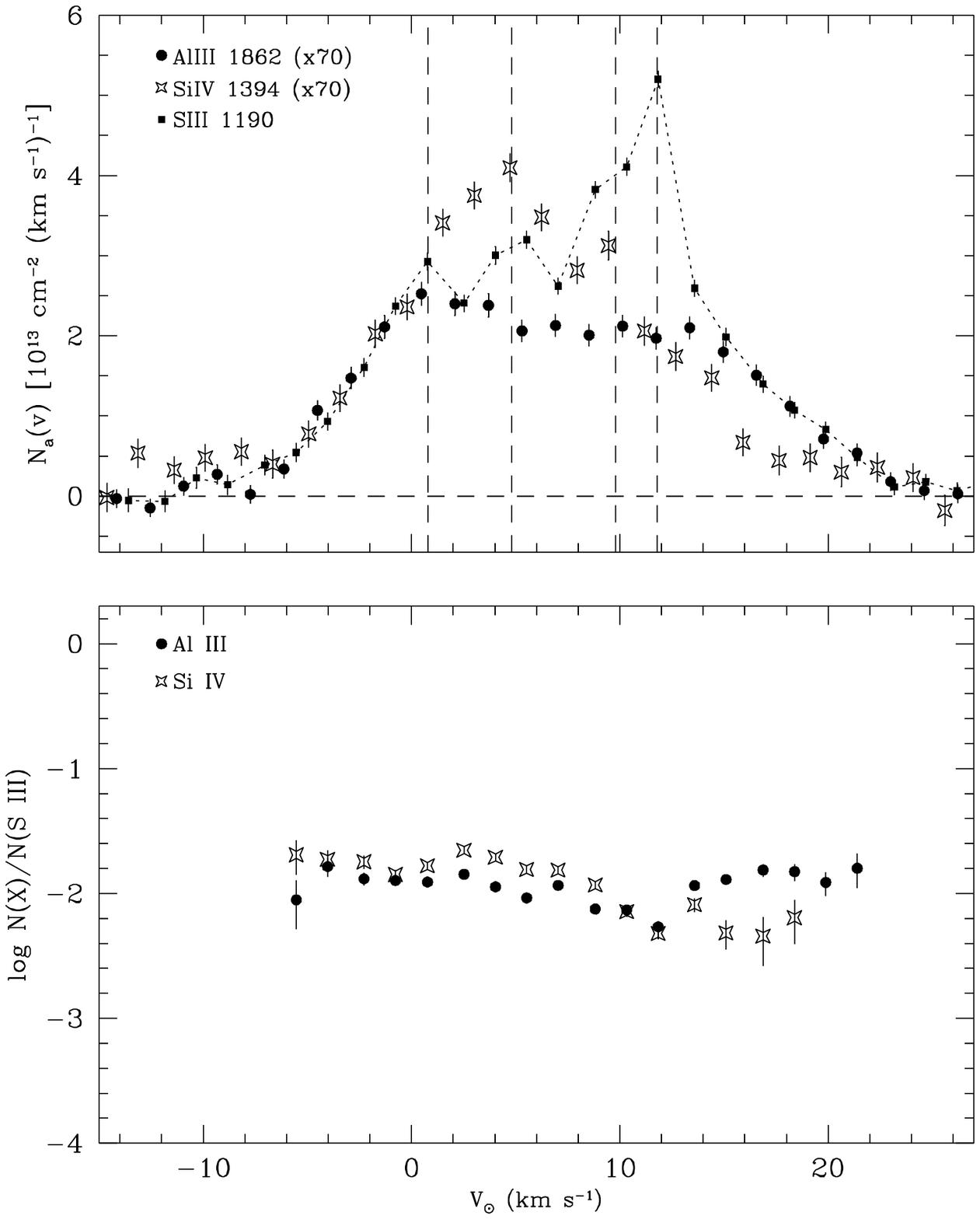} 
\figcaption{The top panel shows the apparent column density profiles
of \protect\al3, \protect\s3, and \protect\ion{Si}{4} observed towards
\protect\xiper\ at 3.5 \protect\kms\ resolution displayed on a
heliocentric velocity scale. The profiles of \protect\al3\ and
\protect\ion{Si}{4} have been scaled upwards by a factor of 70.  The
bottom panel shows the logarithm of the ratio of \protect\al3\ and
\protect\ion{Si}{4} to \protect\s3\ as a function of velocity.  The
vertical dotted lines of the top panel represent the velocities of
\protect\Ha\ emission detected by Reynolds (1988b), though shifted by
$-2.2$ \protect\kms.  \label{fig:xiper}}
\end{figure}

It is also known that the gas-phase abundances of elements increase
with decreasing $\langle n_{\rm H} \rangle \equiv N(\mbox{\HI})/d$ in
the warm neutral ISM (Jenkins 1987; Savage \& Bohlin 1979).  Both the
ionized and neutral gas densities are thought to decrease
exponentially with $z$-height (Dickey \& Lockman 1990; Reynolds 1989),
suggesting that the behavior seen in Figure \ref{fig:deplz} may be
tracing the density-dependence of the gas.  In Figure
\ref{fig:depldens} we plot the values [Al/S]$_i$ versus the electron
density (top) and the average line of sight neutral density (bottom),
as given in Table \ref{table:stars}.  Given the many definitions of
the electron densities (average versus rms, etc.) we have used
different symbols to represent the determinations of rms and average
electron densities (see \S \ref{sec:velocities}).

Figure \ref{fig:depldens} shows a striking relationship between the
average densities and the gas-phase abundances of [Al/S]$_i$ along the
sightlines considered here.  We find that the gas-phase abundance
[Al/S]$_i$ increases with decreasing electron densities.  The observed
relationship between [Al/S]$_i$ and $n_e$ is similar to the observed
dependence of WNM abundances on average sightline neutral hydrogen
(\HI+H$_2$) density (Edgar \& Savage 1989; Jenkins 1987; Savage \&
Bohlin 1979).  The slope of [Al/S]$_i$ versus $\log n_e$ is $-0.37$,
similar to the value $-0.38$ derived for [Fe/H] versus $\log \langle
n_{\rm H} \rangle$ by Jenkins (1987).

Though we are measuring the gas-phase abundance [Al/S]$_i$, i.e., the
abundance of Al to S in the {\em ionized} gas, the bottom panel of
Figure \ref{fig:depldens} shows a significant correlation between
[Al/S]$_i$ and $\log \langle n_{\rm H} \rangle$, the average line of
sight {\em neutral} density.  Savage \etal\ (1990) have also noted a
correlation of $\log N(\mbox{\al3}) / N(\mbox{\HI})$ with decreasing
$\log \langle n_{\rm H} \rangle$.  These authors point out that this
trend may be due to the changing ionization fraction of Al$^{+2}$ with
density, to changing values [Al/H]$_i$ with density, or perhaps both.
Figure \ref{fig:depldens} shows that at least part of the trend
observed by Savage \etal\ is due to the changes in the gas-phase
abundance of Al with average neutral density.  The trend observed by
these authors, which is nicely matched by our data, and that seen in
Figure \ref{fig:depldens} suggest that $\langle n_{\rm H} \rangle$ is
a good indicator of the conditions in the ionized gas.  The slope of
[Al/S]$_i \approx \,$[Al/H]$_i$ versus $\log \langle n_{\rm H}
\rangle$ in our data is less steep than the slope of $\log
N(\mbox{\al3}) / N(\mbox{\HI})$ versus $\log \langle n_{\rm H}
\rangle$ in the Savage \etal\ dataset, suggesting that a combination
of changing gas-phase abundances and ionization fraction is causing
the trend observed by Savage \etal\ and that perhaps the behavior seen
in Figure \ref{fig:depldens} is more widespread than our six
sightlines.

In general Figure \ref{fig:depldens} implies that the ionized and
neutral densities along a sightline follow the same trends, i.e., low
$\langle n_{\rm H} \rangle$ also implies low values of $n_e$.  This
relationship may simply be a manifestation of the known decrease in
both ionized and neutral gas densities as a function of height above
the Galactic plane.  This is also the expected behavior if neutral
clouds with ionized edges are providing most of the observed
absorption (though see Reynolds \etal\ 1998a).  Spitzer (1985) and
Jenkins, Savage, \& Spitzer (1986) have interpreted the dependence of
elemental gas-phase abundances on the average sightline density
$\langle n_{\rm H} \rangle$ in the neutral ISM as a dependence on the
relative contribution of two neutral media: clouds (cold and warm) and
an intercloud medium.  In this picture the warm intercloud medium has
greater gas-phase abundances than the denser clouds.  With an
appropriate mix of each medium, the integrated gas-phase abundance for
a given line of sight can be reproduced.  Perhaps a variation on this
scenario is also appropriate for the ionized medium of our Galaxy.
The relative mix of clouds may be similar between the neutral and
ionized phases depending on the poorly-known relationship between
these phases.

As discussed in \S \ref{subsec:icfwim}, if the WIM of the Galaxy is
photoionized by starlight from OB stars (DM94; Reynolds 1984), our
measurements of \al3\ and \s3\ absorption towards \hd18100\ and
\rholeo\ imply the existence of dust in this important phase of the
ISM.  The values [Al/S]$_i$ for \rholeo\ and \hd18100\ are
significantly higher than those for the other four stars (see Figures
\ref{fig:deplz} and \ref{fig:depldens}), suggesting that the grain
population in the high-latitude WIM has undergone a greater degree of
processing (e.g., by shocks) than have the grains in the low-$z$ \HII\
regions.  For the halo WIM we find a dust phase abundance of $10^6
({\rm Al / H})_d \approx 1.9$ to 2.1.  It would appear that
$\sim20\%-30\%$ of the Al has been liberated from the solid phase in
the high-$z$ WIM compared with the results derived above for the disk
\HII\ regions.

Very little is known about dust in the WIM of the Milky Way (or other
galaxies) from previous studies.  Though in principle one might be
able to separate the thermal dust emission in the WIM from that of the
\HI\ and H$_2$ gas, this has proven difficult.  Boulanger \etal\
(1996) have studied the correlation of the far infrared (FIR) flux
with the column density of neutral hydrogen $N(\mbox{\HI})$ at high
Galactic latitudes.  They find that the correlation between the
$\lambda$21-cm \HI\ emission and the FIR emission detected by the
Cosmic Background Explorer mission is quite good.  They are not,
however, able to rule out a dust abundance in the WIM similar to that
observed in the neutral component.

The existence of dust is important for maintaining the temperature of
the WIM (Reynolds \& Cox 1992; Dettmar \& Schulz 1992).  Reynolds \&
Cox (1992) show that the heating of the WIM may be in large part
provided by photoelectron emission from grains.  This requires the
amount of grain heating per H atom to be similar to that found in the
WNM.  The presence of grains in the WIM has important ramifications
for the diagnostic emission lines used to study this gas.  Reynolds \&
Cox point out that the total heating per H nucleus per second in the
low-density WIM may be twice that of a typical, higher-density \HII\
region.  This has profound effects on the forbidden lines that provide
the cooling for the gas.  The increased cooling required over \HII\
region gas increases the ratios of [\ion{S}{2}], [\ion{N}{2}], and
[\ion{O}{3}] to \Ha\ over \HII\ regions.  Reynolds \& Cox suggest this
extra heating may in part be responsible for the enhanced forbidden
line strengths observed from the WIM of our Galaxy and others
(Reynolds 1985; Rand 1997).

Our data suggest that the destruction of the dust grains in the
high-$z$ WIM has not been extreme.  The value $\sim20\%-30\%$ given
above for the amount of Al liberated from the solid- to gas-phase when
going from disk \HII\ regions to the halo WIM is consistent with
differential measurements of warm neutral cloud abundances between the
disk and the halo.  For a small sample of halo and warm disk clouds,
Howk \etal\ (1998) find roughly $20\% - 30\%$ of the dust-phase Fe and
Si on average have been returned to the gas-phase between the disk and
halo clouds.  Therefore, the processing of grains in the halo WIM does
not appear to be significantly greater than that experienced by clouds
associated with the halo WNM.

This discussion does not suggest that grain destruction mechanisms
have not played an important role in the evolution of the gas being
considered.  The $\sim20\%-30\%$ of dust-phase Al we see returned to
the gas-phase in the halo WIM may be material that was initially bound
in a refractory coating or mantle surrounding the grains.  Savage \&
Sembach (1996b, see their Table 7) have tabulated the dust-phase
abundances of a number of elements.  The derived dust-phase abundances
for the WNM of the Galactic disk give the composition of the grain
cores and mantles, while the observed abundances of the halo material
gives information on the composition of the resilient grain cores that
have probably been stripped of their mantles (Sembach \& Savage 1996).
Savage \& Sembach (1996b) argue that the Fe returned to the gas-phase
in halo material comes predominantly from the mantles thought to
surround the resilient cores that survive the trip into the halo.
They find the dust-phase abundance of Fe in grain cores to be $10^6
({\rm Fe/H})_d = 25$, while for the mantles they find $10^6 ({\rm
Fe/H})_d = 7$.  Thus $\sim22\%$ of the Fe incorporated into grains in
the Milky Way disk resides in a mantle that is relatively easily
stripped.  The liberated Al seen in the WIM at high-$z$ may also come
from the mantles of grains, leaving the resilient grain cores to
account for the remaining $70\%-80\%$ of the Al missing from the gas
phase.

In considering the multiphase structure of neutral clouds in the
Galactic halo, Wolfire \etal\ (1995b; see also Wolfire \etal\ 1995a)
show that the stability of multiphase neutral clouds is affected by
the intrinsic abundances in the gas and by the dust content of the
clouds.  Thus, the presence of dust in ionized halo clouds will have
important implications for the physical structure of the resulting
neutral clouds if the ionized gas recombines.  Indeed, in the case
where the dusty multiphase clouds envisioned by Wolfire \etal\ (1995b)
are situated above the disk of the galaxy, where they are bathed in
ionizing radiation from the disk (assuming photons are able to leak
out of the disk), they will be surrounded by ionized skins.  If no
processes beyond photoionization are responsible for producing the
ionized edges of such clouds, the differences in the dust content of
the neutral and ionized phases should be minimal.  This may explain
why the gas-phase abundances in the ionized gas towards \rholeo\ and
\hd18100, where the neutral and ionized phases of the ISM seem to
coexist (at least in velocity space), are so similar to the derived
refractory element gas-phase abundances in warm neutral halo clouds
(Sembach \& Savage 1996).

\section{SUMMARY}
\label{sec:summary}

This work represents one of the first absorption line studies of the
WIM of the Galaxy.  The observations imply the existence of Al- and
Fe-bearing dust grains in the ionized gas of the Galactic disk and
halo.

A summary of the work presented here and our major conclusions is as
follows:

1) We present archival \ghrs\ intermediate- and high-resolution
 absorption line observations of the moderately-ionized species \al3\
 and \s3\ in the ionized ISM towards six stars.  The sightlines
 towards \zoph, \xiper, \betasco, and \mucol\ probe primarily \HII\
 region gas in the Galactic disk.  The extended high-latitude
 sightlines towards \hd18100\ and \rholeo\ probe the WIM of the Galaxy
 at high $z$.  Results for \fe3\ from the literature are presented for
 \mucol, \betasco, and \zoph.

2) We show, with the possible exceptions of the sightlines to
   \betasco\ and \hd18100, that collisional ionization does not likely
   explain the origin of the observed amounts of \al3\ and \s3.

3) We have computed a grid of photoionization equilibrium models for
 low-density regions excited by late-O/early-B stars using the
 \cloudy\ code (Ferland \etal\ 1998).  We show using our
 photoionization models that the ionization corrections for
 determining [Al/S]$_i$, [P/S]$_i$, and [Si/S]$_i$ using the species
 \al3, \p3, \si3, and \s3 are relatively insensitive to the ionization
 parameter and the effective temperature of the ionizing spectrum.
 Deriving [Fe/S]$_i$ from the ratio $N(\mbox{\fe3})/N(\mbox{\s3})$
 requires a greater knowledge of the stellar effective temperature.

4) We derive the logarithmic gas-phase abundances [Al/S]$_i \approx
 [{\rm Al/H}]_i$ in the ionized material towards the six stars in our
 sample using the results of our photoionization modelling.  All of
 these stars have [Al/S]$_i$ ranging from $-1.2$ to $-0.4$.  Since S
 is normally not incorporated into dust, these abundance results
 indicate the incorporation of Al into dust grains in the ionized
 material along these six sightlines, though for the most distant
 stars, we cannot rule out the confusing effects of collisional
 ionization.  For three stars we find [Fe/S]$_i$ ranges from $-1.6$ to
 $-0.9$.

5) The gas-phase abundances [Al/S]$_i$ and [Fe/S]$_i$ derived here for
 the disk sightlines probing \HII\ region material show that a
 significant degree of Al incorporation into grains is still present
 in the vicinity of stars, i.e., neither the UV radiation fields from
 the stars nor any shocks associated with the stellar winds from these
 stars are sufficient to completely disrupt the refractory grains.
 The abundances we derive here are similar to the gas-phase Fe
 abundances [Fe/H]$_i$ derived for the Orion nebula using emission
 lines (e.g., Osterbrock \etal\ 1992; Baldwin \etal\ 1996).

6) If the WIM of the Galaxy is ionized by the light from OB stars
 (Reynolds 1984), the observed \al3\ and \s3\ column densities towards
 \hd18100\ and \rholeo\ imply the existence of Al-bearing dust in the
 WIM.  To our knowledge this is the first evidence for dust in the
 Galactic diffuse ionized gas, though the effects of collisional
 ionization cannot be ruled out, particularly for the sightline
 towards \hd18100.

7) The gas-phase abundances [Al/S]$_i$ in the ionized material
 increases with height $z$ above the Galactic plane.  Further, the
 values [Al/S]$_i$ increase with decreasing average or rms electron
 densities {\em and} with decreasing average sightline neutral
 hydrogen density.  This behavior is similar to that of the gas-phase
 refractory abundances in the warm neutral medium of the galaxy
 (Jenkins 1987; Bohlin \& Savage 1979).  This general trend implies a
 greater return of elements to the gas phase in more diffuse
 environments.

8) The observed values of [Al/S]$_i$ are similar to the abundances of
 other refractory elements seen in the WNM of the disk and halo.
 Further, the variation of [Al/S]$_i$ with density (electron or
 neutral) is also similar to the values observed for refractory
 elements in the WNM (esp., Fe or Mn).  Our analysis implies that the
 processing of dust grains in the ionized gas may not be much
 different than that in the low-density warm neutral medium.

9) We discuss the velocity structure of the WIM along the two
 high-latitude sightlines in our sample.  These directions show a very
 close relationship between the tracers of neutral material and
 ionized gas, similar to the correspondence observed by Spitzer \&
 Fitzpatrick (1993).  The data show no kinematic evidence for a
 separation of the ionized and neutral material.  This is consistent
 with a partially-ionized medium in which the neutrals and ions are
 well mixed (e.g., SF93), the neutral clouds are surrounded by ionized
 envelopes (e.g., McKee \& Ostriker 1977), or other more complex
 scenarios.

\acknowledgements

We have benefitted from conversations with R. Reynolds, B. Benjamin,
and J. Mathis in preparing this manuscript.  We also thank E. Jenkins
for a careful reading of our work and for suggestions that improved
the final result.  We thank E. Fitzpatrick for sharing his component
fitting software with us, and M. Haffner, R. Reynolds, and the \wham\
group for sharing their results in advance of publication.  We feel
G. Ferland and his co-workers at the University of Kentucky have done
a great service for the astronomical community in making available
their photoionization code CLOUDY.  This work has only proceeded
because of our confidence in the implementation of the physics
included in CLOUDY.  [The CLOUDY code and documentation, e.g., Ferland
(1996), can be accessed via the address {\tt
http://www.pa.uky.edu/$\sim$gary/cloudy/} on the World Wide Web.]  The
archival \ghrs\ data used here were acquired from the \ghrs\
Instrument Definition Team archive ({\tt
http://hrssun.gsfc.nasa.gov/}).  This research has made use of the
SIMBAD database, operated at CDS, Strasbourg, France.  JCH recognizes
support from a NASA Graduate Student Researcher Fellowship under grant
number NGT-5-50121.

{\small
\begin{planotable}{lcccccccr}
\tablewidth{0pc}
\tablenum{1}
\tablecolumns{9}
\tableheadfrac{0.15}
\tablecaption{UV Probes of Weakly Ionized Gas \label{table:uvprobes}}
\tablehead{
\colhead{ } & \colhead{ } & \colhead{ } & \colhead{IP} & 
\colhead{IP} & \colhead{ } & \colhead{ } & \colhead{ } &
\colhead{ } \\
\colhead{Ion} & 
\colhead{$\lambda$\tablenotemark{a}} & 
\colhead{$f$\tablenotemark{b}} & 
\colhead{$({\scriptstyle X^{i-1} \rightarrow X^i})$} & 
\colhead{$({\scriptstyle X^i \rightarrow X^{i+1}})$} &
\colhead{$\log\{X/{\rm H}\}_\odot$\tablenotemark{c}} & 
\colhead{[$X$/H]$_{halo}$\tablenotemark{d}} & 
\colhead{$\log N(X)_{\rm H\,II}$\tablenotemark{e}} &
\colhead{$\tau_o$\tablenotemark{f}} \\
\colhead{ } & \colhead{[\AA]} & \colhead{ }   &
\colhead{[eV]} & \colhead{[eV]} & \colhead{$+12.00$} &
\colhead{ } & \colhead{[cm$^{-2}$]} & \colhead{ } }
\startdata
\ion{C}{3} & 977.020 & 0.762 & 
	24.4 & 47.9 & 8.55 &
	$-$0.4\tablenotemark{g}\phn & 14.46 & 86.4 \nl
\ion{N}{2} & 1083.990 & 0.103 &
	14.5 & 29.6 & 7.97 &
	\phs0.0\phn  & 14.88 & 36.9 \nl
\ion{N}{3} & 989.799 & 0.107 &
	29.6 & 47.4 & 7.97 &
	\phs0.0\phn & 14.11 & 5.95 \nl
\ion{Si}{3} & 1206.500 & 1.67  &
	16.3 & 33.5 & 7.55 &
	$-$0.25 & 13.96 & 111.7 \nl
\ion{Fe}{3} & 1122.526 & 0.0788 & 
	16.2 & 30.7 & 7.51 &
	$-$0.6\phn & 13.78 & 4.6 \nl
\ion{S}{3} & 1190.208 & 0.0222 & 
	23.3 & 34.8 & 7.27 &
	\phs0.0\phn & 14.02 & 1.81 \nl
 	& 1012.502 & 0.0355 & 
	23.3 & 34.8 & 7.27 &
	\phs0.0\phn & 14.02 & 2.46 \nl
\ion{Ar}{2} & 919.781 & 0.00887 &
	15.8 & 27.6 & 6.56 &
	\phs0.0\tablenotemark{h}\phn 
	& 13.37 & 0.14 \nl
\ion{Al}{3} & 1862.789 & 0.279 & 
	18.8 & 28.4 & 6.48 &
	$-$0.6\tablenotemark{i}\phn
		& 12.42 & 0.82 \nl
	    & 1854.716 & 0.560 &
	18.8 & 28.4 & 6.48 &
	$-$0.6\tablenotemark{i}\phn 
		& 12.42 & 1.65  \nl 
\ion{Cr}{3} & 1040.050 & 0.122 &
	16.5 & 31.0 & 5.68 &
	$-$0.35 & 12.21 & 0.17 \nl
	    & 1033.331 & 0.0640 &
	16.5 & 31.0 & 5.68 &
		$-$0.35 & 12.21 & 0.09 \nl
	    & 1030.100 & 0.0625 &
	16.5 & 31.0 & 5.68 &
		$-$0.35 & 12.21 & 0.09 \nl
\ion{P}{3}  & 1334.813    & 0.0253 &
	19.7 & 30.2 & 5.57 &
	\phs0.0\phn & 12.30 & 0.04 \nl
	    & \phn998.000 & 0.112  & 
	19.7 & 30.2 & 5.57 &
	\phs0.0\phn & 12.30 & 0.14 \nl
\ion{Ti}{3} & 1298.697 & 0.0951 & 
	13.6 & 27.5 & 4.93 &
	$-$0.65 & 11.28 & 0.02 \nl
	    & 1295.884 & 0.0668 &
	13.6 & 27.5 & 4.93 &
	$-$0.65 & 11.28 & 0.01 \nl
\enddata
\tablenotetext{a}{ Vacuum wavelengths from Morton (1991).}
\tablenotetext{b}{ Oscillator strengths from Morton (1991).} 
\tablenotetext{c}{ The logarithmic ``solar'' abundances of the
	elements, $\log \{X/{\rm H} \}_\odot$.  We have adopted the
	solar system meteoritic abundances from Anders \& Grevessee
	(1989) except for C which is the photospheric value from
	Grevesse \& Noels (1993).}
\tablenotetext{d}{ Typical values of the logarithmic
	normalized gas-phase abundance seen in warm neutral halo
	clouds, defined such that [$X$/H]$\, = \log \{N(X)/N({\rm
	H})\} - \log \{ X/{\rm H} \}_\odot$.  These values are for the
	warm cloud at $v_\odot \approx +41$ km s$^{-1}$ seen towards
	$\mu$~Col (Sofia, Cardelli, \& Savage 1993; Shull \& York
	1977), which is typical of ``halo''-type neutral clouds (Howk
	{\em et al.}  1998).}
\tablenotetext{e}{ Expected column density of each species for a
	fully-ionized cloud with $\log N({\rm H\,II}) = 19.0$  
	cm$^{-2}$ as calculated using the CLOUDY photoionization
	equilibrium code (see \S 4.1).  The assumed gas-phase
	abundances of each of the elements are dictated by the solar
	system values modified by the normalized gas-phase abundances
	typical of halo material.  The shape of the ionizing
	spectrum was taken to be that of an O9.5V star with $T_{eff} =
	33,000$ K.}
\tablenotetext{f}{ Expected peak optical depth of each line
	assuming the column densities given in column 8 with
	$b$-values appropriate for gas at $T=10,000$ K and no
	non-thermal broadening.}
\tablenotetext{g}{ The gas-phase abundance of C has only
	been measured accurately in cool neutral clouds found in the
	disk.  We tentatively adopt the Cardelli {\em et al.} average
	[C/H]$\, \approx -0.4$ for our ``halo'' abundances.}
\tablenotetext{h}{ The ratio Ar$\,$I/H$\,$I has
	recently been measured to be significantly sub-solar along a
	number of low-$N({\rm H\,I})$, partially ionized sightlines by
	Sofia \& Jenkins (1998).  However, these authors argue that
	the large ionization cross-section of Ar$^{\rm o}$ implies
	that much of the Ar may reside in the form of Ar$^{+}$, which
	they did not observe.  We therefore adopt [Ar/H]$\, \approx
	0.0$ in warm neutral halo gas.}
\tablenotetext{i}{ The abundance of Al is poorly known in the warm
	neutral medium because the Al$\,$II $\lambda$1670 \AA\ line
	requires very large saturation corrections (Barker \etal\
	1984).  We have assumed [Al/H]$\, \approx\,$[Fe/H].}
\end{planotable}
}

\pagebreak

\begin{planotable}{lcccclcccc}
\tablewidth{0pc} 
\tablecolumns{10} 
\tablecaption{Stellar and Sightline Properties\label{table:stars}}
\tablehead{
\colhead{Star} & 
\colhead{$l$}  & \colhead{$b$}      &
\colhead{$d$\tablenotemark{a}}      &
\colhead{$z$\tablenotemark{b}}      &
\colhead{Spectral}                  &
\colhead{$T_{eff}$\tablenotemark{c}}&
\colhead{E(B-V)\tablenotemark{d}} &
\colhead{$\langle n({\rm H}) \rangle$\tablenotemark{e}} &
\colhead{$n_e$\tablenotemark{f}}\\
\colhead{HD(Name)}      & \colhead{}      & \colhead{} &
\colhead{[pc]}  & \colhead{[pc]} &
\colhead{Type}  & \colhead{[K]} &
\colhead{[mag]} &
\colhead{[cm$^{-3}$]} &
\colhead{[cm$^{-3}$]} 
}
\startdata 
24912($\xi$ Per) & $160\fdg4$ & $-13\fdg1$ & 540 
	& 122 & O7.5 I & 36,000 
	& 0.33 & 1.2\phn & 1.4\phn \nl
38666($\mu$ Col) & $237\fdg3$ & $-27\fdg1$ & 400 
	& 180 & O9.5 V & 33,000 
	& 0.01 & 0.06 & 0.2\phn \nl
149757($\zeta$ Oph) & \phn \phn$6\fdg3$   & $+23\fdg6$  & 140
	& 56  & O9.5 V & 31,900 
	& 0.32 & 3.1\phn & 4.0\phn \nl
144217($\beta^1$ Sco) & $353\fdg2$ & $+23\fdg6$ & 160 
	& 60 & B0.5 V  & 28,000 
	& 0.20 & 2.9\phn & \nodata \nl
91316($\rho$ Leo)      & $234\fdg9$ & $+52\fdg8$ & 870 & 690 &
		B1 Ib & 26,800 
		& 0.05 & 0.10 & 0.07 \nl
18100	& $217\fdg9$ & $-62\fdg7$ & 3100 & 2800 &
		B1 V  & 26,400 
		& 0.02 & 0.01 & 0.07 \nl
\enddata
\tablenotetext{a}{Distances, with the exception of $\rho$ Leo and HD
	18100, are based upon {\em Hipparcos} measurements of stellar
	parallax (Perryman {\em et al.} 1997).  The distance estimates
	to $\rho$ Leo and HD 18100 are from Keenan, Brown, \& Lennon
	(1986) and Diplas \& Savage (1994), respectively.}
\tablenotetext{b}{Distance from the mid-plane of the galaxy given the
	derived distances.}
\tablenotetext{c}{Adopted effective temperatures for each of the
	stars.  These data are from the following: Code {\em et al.}
	(1976; $\zeta$ Oph); Holmgren {\em et al.}
	(1997; $\beta^1$ Sco); Howarth \& Prinja (1989; $\mu$ Col);
	Keenan {\em et al.} (1986; HD 18100); Keenan \& Dufton
	(1983; $\rho$ Leo); Sokolov (1995; 23 Ori); and
	Vacca, Garmany, \& Shull (1996; $\xi$ Per).} 
%
%
\tablenotetext{d}{The dust color excess, E(B-V), is taken from Diplas
	\& Savage (1994).}
\tablenotetext{e}{Average line of sight neutral hydrogen densities
	towards these stars, where $\langle n({\rm H}) ~ \rangle
	\equiv ~ \{ N(\mbox{\protect\ion{H}{1}}) + 2N({\rm H_2}) \} / d$.  We
	have taken values for $N(\mbox{\protect\ion{H}{1}})$ from Diplas \&
	Savage (1994) and $N({\rm H_2})$ from Bohlin, Savage, \& Drake
	(1978).  The value listed for \hd18100\ is $\langle
	n(\mbox{\HI}) \rangle$.}
\tablenotetext{f}{Electron densities along these lines of sight, when
	available.  The data presented here are from Howk {\em et al.}
	(1998; $\mu$~Col); Reynolds (1988b; $\xi$~Per and
	$\zeta$~Oph); Savage \& Sembach (1996; HD~18100); and this
	work ($\rho$ Leo).  The values quoted for $\xi$~Per and
	$\zeta$~Oph are the values $\langle n_e^2 \rangle ^{1/2}$
	characteristic of the \protect\ion{H}{2} regions surrounding these
	stars.  For \mucol\ the value quoted is $\langle n_e \rangle$
	of the \HII\ region.  The values quoted for HD~18100 and
	$\rho$ Leo are the values $\langle n_e \rangle$ of the warm
	ionized gas in these directions.}
\end{planotable}

\pagebreak

\begin{planotable}{llccccc}
\tablewidth{0pc} 
\tablecolumns{7} 
\tablecaption{Log of GHRS Archival Data\label{table:log}}
\tablehead{
\colhead{Star} &
	\colhead{Spectral Range} &
        \colhead{Rootname\tablenotemark{a}} & 
        \colhead{Exp.\tablenotemark{b}}  & 
        \colhead{Mode} & 
	\colhead{Aper.\tablenotemark{d}} &
	\colhead{\small FP-SPLIT/} \\
	& \colhead{[\AA]} & 
	& \colhead{[sec]} 
	& \colhead{\& Order\tablenotemark{c}} 
	& \colhead{} 
	& \colhead{\small COSTAR?\tablenotemark{e}}}
\startdata 
$\xi$ Per 
& 1187.7-1194.2 & Z0GY010LT & 691.2 & Ech-A/47 & SSA & 4/F \nl
(HD 24912)
& 1857.1-1867.4	& Z0GY011ST & 172.8 & Ech-B/30 & SSA & 4/F \nl
\nl
$\mu$ Col 
& 1184.8-1191.1 & Z2AF010PT & 108.8 & Ech-A/47 & LSA & 0/T \nl
(HD 38666)
& 1184.8-1191.1 & Z2C0020PP & 108.8 & Ech-A/47 & LSA & 0/T \nl
& 1846.7-1856.9 & Z2D40118T & 54.4 & Ech-B/30 & LSA & 0/T \nl
& 1859.8-1869.8 & Z2CX010LT & 27.2 & Ech-B/30 & LSA & 0/T \nl
& 1855.9-1866.0 & Z2D4020KT & 54.4 & Ech-B/30 & LSA & 0/T \nl
\nl
$\zeta$ Oph  
& 1188.6-1195.0 & Z2VX010CT & 691.2 & Ech-A/47 & SSA & 4$^*$/T \nl
(HD 149757)
& 1189.8-1196.2 & Z2VX010ET & 691.2 & Ech-A/47 & SSA & 4$^*$/T \nl
& 1856.9-1866.9 & Z0LD020TT & 172.8 & Ech-B/30 & SSA & 4$^*$/F \nl
\nl
$\beta^1$ Sco 
& 1180.1-1216.3 & Z0YU010AT & 172.8 & G160M/01 & SSA & 4/F \nl
(HD 144217)
& 1856.7-1866.7 & Z0YU020AT & 172.8 & Ech-B/30 & SSA &  4$^*$/F \nl
 \nl
$\rho$ Leo  
& 1188.7-1195.3 & Z2ZX010CT & 870.4 & Ech-A/47& SSA & 4/T \nl
(HD 91316)
& 1852.7-1862.9 & Z0ZI0314T & 86.4 & Ech-B/30 & SSA & 0/F \nl
& 1853.5-1863.6	& Z0ZI0315T & 86.4 & Ech-B/30 & SSA & 0/F \nl
& 1854.3-1864.4 & Z0ZI0316T & 86.4 & Ech-B/30 & SSA & 0/F \nl
& 1855.0-1865.1 & Z0ZI0317T & 86.4 & Ech-B/30 & SSA & 0/F \nl
 \nl
HD 18100
& 1181.4-1217.6 & Z13Z010AT & 1324.8 & G160M/01 & SSA & 4/F \nl
& 1842.9-1876.9 & Z13Z010NM & 1209.6 & G160M/01 & SSA & 4/F \nl
\enddata
\tablenotetext{a}{STScI archival rootname.}
\tablenotetext{b}{Total exposure time given in seconds.} 
\tablenotetext{c}{Grating mode and spectral order used for the
        observation.} 
\tablenotetext{d}{Aperture used for the observation.  The LSA subtends
        $1\farcs 74 \times 1\farcs 74$ on the sky for post-COSTAR
        observations, $2\farcs 0 \times 2\farcs 0$ for pre-COSTAR
        data.  The pre-COSTAR SSA subtends $0\farcs 25 \times 0\farcs
        25$, while the post-COSTAR SSA is $0\farcs 22 \times 0\farcs
        22$ on the sky.}
\tablenotetext{e}{The number of FP-SPLIT sub-exposures composing each
	observation.  Asterisks mark those observations for which we
	have explicitly derived the fixed-pattern noise spectrum and
	removed it. This column also notes with a ``T'' those
	observations taken after the installation of COSTAR, and with
	an ``F'' for those taken before
        COSTAR.}
\end{planotable}

\pagebreak

\begin{planotable}{lccccccc}
\tablewidth{7in} 
\tablecolumns{8} 
\tablecaption{Equivalent Widths and Column Densities of \protect\ion{S}{3},
\protect\ion{Al}{3} and \protect\ion{Fe}{3}\label{table:columns}}
\tablehead{
\colhead{} & 
\multicolumn{3}{c}{ $W_\lambda \, \pm \sigma$ [m\AA]
\tablenotemark{a}} & \colhead{} &
\multicolumn{3}{c}{ $\log N \, \pm \sigma$ [cm $^{-2}$]
\tablenotemark{b}}\\
\cline{2-4}  \cline{6-8}
 \colhead{Star} &
\colhead{\protect\ion{S}{3}}   &
 \colhead{\protect\ion{Al}{3}} &
 \colhead{\protect\ion{Al}{3}} &
	\colhead{} &
\colhead{\protect\ion{S}{3}}   &  \colhead{\protect\ion{Al}{3}} &
\colhead{\protect\ion{Fe}{3}\tablenotemark{c}}
\\
 \colhead{} &  \colhead{$\lambda$1190.2 \AA} &  
 \colhead{$\lambda$1854.7 \AA} &
 \colhead{$\lambda$1862.8 \AA} & 
 \colhead{} & \colhead{} & \colhead{} 
}
\startdata 
$\xi$ Per	&
	$87\pm3$ & \nodata & $51.2\pm1.7$  & &
	$14.82\pm0.02$ & $12.85\pm0.02$  & \nodata \nl
$\mu$ Col	&
	$16.3\pm0.6$ & $15.0\pm2.0$ & \phn$9.9\pm1.5$ & &
	$13.82\pm0.02$ & $12.01\pm 0.05$ & 
	$13.37 \ ^{+ \, 0.09}_{- \, 0.11}$\phn \nl
$\zeta$ Oph	&
	$66.6\pm1.4$   & \nodata & $19.7\pm0.8$ & &
	$14.76\pm0.02$ & $12.42\pm0.02$ & $13.45\pm0.10$ \nl
$\beta^1$ Sco	&
	$27\pm5$\tablenotemark{d} & \nodata & \phn$9.5\pm0.7$ & &
	$13.98\pm0.08$\tablenotemark{d} & $12.06\pm0.04$ & $13.10\pm0.10$ \nl
$\rho$ Leo & 
	$12.7\pm0.8$   & $17.5\pm1.6$\tablenotemark{e} & \nodata & &
	$13.72\pm0.03$ & $12.06\pm0.04$\tablenotemark{e} & \nodata \nl
HD 18100   &	
	$54\pm8$\tablenotemark{d}        & $74\pm6$  & $40\pm6$ & & 
	$14.29\pm0.06$\tablenotemark{d}  & $12.70\pm0.04$ & \nodata \nl
\enddata
\tablenotetext{a}{Equivalent widths $W_\lambda$ in m\AA\ for the lines
	of \protect\ion{S}{3} and \protect\ion{Al}{3} with $1\sigma$ uncertainties.}
\tablenotetext{b}{Column densities of interstellar \protect\ion{S}{3} and
	\protect\ion{Al}{3}  in	units atoms cm$^{-2}$.  Also given are the
	$1\sigma$ error estimates for these measurements.} 
\tablenotetext{c}{The \protect\ion{Fe}{3} column densities quoted here are
	taken from the following: Howk {\em et al.} (1998; $\mu$ Col);
	Morton (1975; $\zeta$ Oph); and Savage \& Bohlin (1979;
	$\beta^1$ Sco).  These column densities are all derived from the
	\protect\ion{Fe}{3} $\lambda 1122.5$ \AA\ line. The latter two are
	based upon {\em Copernicus} observations, while the former
	comes from GHRS G160M observations.  The {\em Copernicus}
	observations have been adjusted by $-0.15$ dex to account for
	newer oscillator strengths (Morton 1991; $f=0.07884$) and the
	errors are estimates by the current authors.  In the case of
	$\zeta$ Oph and possibly $\beta^1$ Sco, \protect\ion{C}{1} absorption
	could be contributing to these column densities.  For \mucol\
	Howk {\em et al.} (1998) have put restrictive limits on the
	degree of this contamination and find it not to be significant
	compared with the quoted uncertainties.}
\tablenotetext{d}{These values are based upon \ghrs\ G160M data and
	have been derived through a component fitting analysis.}
\tablenotetext{e}{These values are for the velocity range $-19
	\leq v_\odot \leq +2$ \kms, which corresponds to the
	uncontaminated velocity range for the \s3\ absorption.  The
	total integrated sightline values  for \al3\ are $W_\lambda =
	29\pm3$ m\AA\ and $\log N(\mbox{\al3}) = 12.27\pm0.05$.}
\end{planotable}

\pagebreak

\begin{planotable}{ccccccccc}
\tablewidth{0pc} 
\tablecolumns{9} 
\tablecaption{CLOUDY \protect\ion{H}{2} Region Model for an O9.5 V
	Star\tablenotemark{a}\label{table:cloudy_density}}  
\tablehead{
\colhead{$n_{\rm H}$\tablenotemark{b}} & 
\colhead{$\log (q)$\tablenotemark{c}} & 
\colhead{$\log (U)$\tablenotemark{d}} & 
\colhead{$\log  x({\rm S}^{+2})$\tablenotemark{e}}  &
\multicolumn{5}{c}{$\log  x(X^i) / x({\rm S}^{+2})$\tablenotemark{f}} \\
\cline{5-9}
[cm$^{-3}$] 
& & & &
\colhead{$ {\rm Al}^{+2} $} &
\colhead{$ {\rm Fe}^{+2}$}  &
\colhead{$ {\rm Si}^{+2}$}  &
\colhead{$ {\rm Si}^{+3}$}  &
\colhead{$ {\rm P}^{+2}$}   
}
\startdata 
0.02  & $-$4.1 & $-4.3$ & $-$0.45 & $-$0.25 &   
	0.24 &  $-$0.15 &  $-$2.03 &   $-$0.01 \nl
0.05  & $-$3.7 & $-4.1$ & $-$0.40 & $-$0.24 &   
	0.21 &  $-$0.13 &  $-$2.08 &  $-$0.01\nl  
0.1   & $-$3.4 & $-4.0$ & $-$0.36 & $-$0.23 &   
	0.19 &  $-$0.12 &  $-$2.14 &  $-$0.02\nl 
0.2   & $-$3.1 & $-3.9$ & $-$0.33 & $-$0.23 &   
	0.17 &  $-$0.11 &  $-$2.22 &  $-$0.02\nl 
0.5   & $-$2.7 & $-3.8$ & $-$0.28 & $-$0.22 &   
	0.14 &  $-$0.10 &  $-$2.36 &  $-$0.02\nl 
1.0   & $-$2.4 & $-3.7$ & $-$0.25 & $-$0.21 &   
	0.12 &  $-$0.09 &  $-$2.46 &  $-$0.02\nl 
1.5   & $-$2.2 & $-3.6$ & $-$0.24 & $-$0.20 &   
	0.11 &  $-$0.09 &  $-$2.55 &  $-$0.02\nl 
10.0  & $-$1.4 & $-3.4$ & $-$0.18 & $-$0.18 &   
	0.08 &  $-$0.07 &  $-$2.94 &  $-$0.02\nl 
100.0 & $-$0.4 & $-3.0$ & $-$0.15 & $-$0.20 &   
	0.10 &  $-$0.06 &  $-$3.77 &  $-$0.03\nl 
\enddata 
\tablenotetext{a}{The ionization fractions reported here are the
radially-averaged values of nebular models with an input stellar
effective temperature $T_{eff} = 33,000$ K and total luminosity $\log
L_* / L_\odot = 4.4$.  The ionizing photon luminosity in these models
is $\approx 4 \times 10^{47}$ photons s$^{-1}$.}
\tablenotetext{b}{Total hydrogen density (neutral plus ionized) used
in the model.}
\tablenotetext{c}{$q \equiv n_{\rm H} f^2 L_{50}$, where $n_{\rm H}$
is the ambient hydrogen density, $f$ the filling factor, and $L_{50}$
the stellar ionizing flux in units of $10^{50}$ photons s$^{-1}$.}
\tablenotetext{d}{$U \equiv L/(4 \pi R_S^2 n_{\rm H} c)$, where
$n_{\rm H}$ is the ambient hydrogen density and $L$ is the hydrogen
ionizing photon luminosity in photons~s$^{-1}$.  The Str\"{o}mgren
radius is $R_S = [ 3L/(4\pi n_{\rm H}^2 f \alpha_B) ]^{1/3}$, where
$\alpha_B$ is the case B recombination coefficient of H.  We give
values assuming $\alpha_B = 2.59\times10^{-13}$ cm$^3$~s$^{-1}$
appropriate for $T_e = 10^4$~K (Osterbrock 1989).}
\tablenotetext{e}{$\log  x(X^i) / x({\rm S}^{+2}) \equiv N({\rm
S}^{+2}) / N({\rm S_{\rm Total}})$.}
\tablenotetext{f}{These columns give $\log  x(X^i) / x({\rm S}^{+2})
\equiv - \log {\rm ICF}$ (see \S \ref{sec:abundances}).}
\end{planotable}

\pagebreak

\begin{planotable}{ccccccc}
\tablewidth{0pc} 
\tablecolumns{7} 
\tablecaption{CLOUDY \protect\ion{H}{2} Region Model for
	$\log (q)=-4.0$\tablenotemark{a}\label{table:cloudy_temp}}    
\tablehead{
\colhead{$T_{eff}$} &
\colhead{$\log x({\rm S}^{+2})$} &
\multicolumn{5}{c}{$\log  x(X^i) / x({\rm S}^{+2})$} \\
\cline{3-7}
[K] & &
\colhead{\phn \phn  \phn$ {\rm Al}^{+2} $} &
\colhead{\phn \phn \phn$ {\rm Fe}^{+2}$}  &
\colhead{\phn \phn \phn$ {\rm Si}^{+2}$}  &
\colhead{\phn \phn \phn$ {\rm Si}^{+3}$}  &
\colhead{\phn \phn \phn$ {\rm P}^{+2}$}   
}
\startdata 
 27,000 &$-0.71$ & $-0.11$ &  \phs$ 0.55$ 
		&  \phs$ 0.12$ & $-4.93$ &  \phs$ 0.12$\nl 
 29,000 &$-0.58$ & $-0.16$ &  \phs$ 0.40$ 
		&  $-0.01$ & $-3.85$ &  \phs$ 0.05$\nl 
 31,000 &$-0.49$ & $-0.21$ &  \phs$ 0.30$ 
		&  $-0.10$ & $-2.78$ &  \phs$ 0.01$\nl 
 33,000 &$-0.44$ & $-0.25$ &  \phs$ 0.23$ 
		&  $-0.14$ & $-2.04$ &  $-0.01$\nl 
 35,000 &$-0.40$ & $-0.30$ &  \phs$ 0.15$ 
		&  $-0.19$ & $-1.58$ &  $-0.04$\nl 
 37,000 &$-0.37$ & $-0.33$ &  \phs$ 0.10$ 
		&  $-0.22$ & $-1.43$ &  $-0.06$\nl
 39,000 &$-0.34$ & $-0.36$ &  \phs$ 0.04$ 
		&  $-0.27$ & $-1.30$ &  $-0.10$\nl 
\nl
\multicolumn{2}{c}{$\log \, \langle {\rm ICF}
		\rangle_{Rad}$\tablenotemark{b}} &  
	$+0.24\pm0.07$ & 
	$-0.17\pm0.19$ & 
	$+0.10\pm0.10$ & 
 	$+2.6\pm1.4$\tablenotemark{c} & 
	$+0.01\pm0.06$ \nl
\multicolumn{2}{c}{$\log \, \langle {\rm ICF}
		\rangle_{Vol}$\tablenotemark{d}} &  
	$+0.37\pm0.07$ & 
	$-0.4\pm0.3$ & 
	$+0.38\pm0.21$ & 
	$+4.3\pm1.4$\tablenotemark{c} & 
	$-0.05\pm0.08$ \nl
\enddata 
\tablenotetext{a}{The ionization fractions and ratios reported here
are the radially-averaged values of nebular models with $\log(q)\equiv
\log(n_{\rm H} f^2 L_{50}) = -4.0$. The alternate definition of the
ionization parameter gives $\log (U) \equiv \log \{ L/(4 \pi R_S^2
n_{\rm H} c) \} = -4.22$ assuming $\alpha_B = 2.59\times10^{-13}$
cm$^3$~s$^{-1}$ appropriate for $T_e = 10^4$ K (Osterbrock 1989).}
\tablenotetext{b}{The mean value of the radially-averaged ionization
corrections and standard deviation about the mean for each species
considered.  These numbers are the average and dispersion of the ICFs
for all stellar effective temperatures considered for models having
$\log(q) =-4.0$ and $-2.0$.}
\tablenotetext{c}{The values of ICF(Si$^{+3}$) are highly uncertain
given the strong dependence upon the stellar effective temperature and
ionization parameter.}
\tablenotetext{d}{The mean value of the volume-averaged ionization
corrections and standard deviation about the mean for each species
considered.  These numbers are the average and dispersion of the ICFs
for all stellar temperatures considered for models having $\log(q)
=-4.0$ and $-2.0$.}
\end{planotable}

\pagebreak

\begin{planotable}{lcccc}
\tablewidth{0pc} 
\tablecolumns{5} 
\tablecaption{Column Density Ratios and Logarithmic Gas-Phase
	Abundances\label{table:ratios}}
\tablehead{
	\colhead{Star} &
	\colhead{$\log N(\mbox{\protect\ion{Al}{3}}) / N(\mbox{\protect\ion{S}{3}})$} &
	\colhead{$\log N(\mbox{\protect\ion{Fe}{3}}) / N(\mbox{\protect\ion{S}{3}})$} &
	\colhead{[Al/S]$_i$} &
	\colhead{[Fe/S]$_i$} 
}
\startdata 
$\xi$ Per 
	& $-1.97\pm0.03$ & \nodata 
	& $-0.87\pm0.08$ & \nodata \nl
$\mu$ Col 
	& $-1.82\pm0.05$ & $-0.40\pm0.11$
	& $-0.78\pm0.08$ & $-0.87\pm0.21$ \nl
$\zeta$ Oph\tablenotemark{a}
	& $-2.23\pm0.03$ & $-1.05\pm0.10$
	& $-1.19\pm0.08$ & $-1.52\pm0.21$ \nl
$\beta^1$ Sco 
	& $-1.92\pm0.09$ & $-0.88\pm0.12$
	& $-1.00\pm0.11$ & $-1.59\pm0.21$ \nl
$\rho$ Leo  
	& $-1.66\pm0.06$ & \nodata 
	& $-0.50\pm0.10$\tablenotemark{b} & \nodata \nl
HD 18100
	& $-1.59\pm0.07$ & \nodata 
	& $-0.43\pm0.10$\tablenotemark{b} & \nodata \nl
\enddata
\tablenotetext{a}{The values given here for $\zeta$~Oph are the for
	the integrated sightline.}
\tablenotetext{b}{The values given here for [Al/S]$_i$ towards
	$\rho$~Leo and HD~18100 assume the shape of the ionizing
	spectrum of the diffuse ionized gas is not very different than
	that of a star with $27,000 \lesssim T_{eff} \lesssim 39,000$
	K.}
\end{planotable}

\pagebreak


\begin{planotable}{lccc}
\tablewidth{0pc}
\tablecolumns{4}
\tablecaption{Measurements of \protect\ion{Si}{4} Column Densities\label{table:si4}}
\tablehead{
\colhead{Star} & \colhead{$\log N(\mbox{\protect\ion{Si}{4}})$}  &
\colhead{$\log [N(\mbox{\protect\ion{Si}{4}})/N(\mbox{\protect\ion{S}{3}})]$} &
\colhead{Ref.}
}
\startdata
$\xi$ Per	&$12.89 \pm0.03$ &	$-1.93$	&	1 \nl
$\mu$ Col	&$12.17 \pm0.05$ &	$-1.65$	&	2 \nl
$\zeta$ Oph	&$12.79 \pm0.02$ &	$-1.97$\tablenotemark{a}
						&	3 \nl
$\beta^1$ Sco	&$12.02 \pm0.03$ &	$-1.96$	&	1 \nl
$\rho$ Leo	&$<11.6$\tablenotemark{b} 
		& $<-2.1$\tablenotemark{b}      &	1 \nl
HD 18100	&$13.10 \pm0.04$ &	$-1.19$	&	4 \nl
\enddata
\tablerefs{(1) This work; (2) Brandt et al (1998); (3) Sembach et al
(1994); (4) Sembach \& Savage (1994).}
\tablenotetext{a}{The value $\log
[N(\mbox{\ion{Si}{4}})/N(\mbox{\ion{S}{3}})]$ given for the
\protect\zoph\ sightline is the value integrated over all velocities.
Figure \ref{fig:zoph} shows that this ratio is a strong function of
velocity.  Near $v_\odot \approx -15$ \kms, corresponding to the peak
of the \protect\ion{Si}{4} absorption, a value of $-1.6$ to $-1.8$ is more
appropriate; while at the peak of the \al3\ absorption near $v_\odot
\approx -8$ \kms, $\log [N(\mbox{\ion{Si}{4}})/N(\mbox{\ion{S}{3}})]
\approx -2.4$ is more appropriate.}
\tablenotetext{b}{This $2 \sigma$ upper limit to $\log
N(\mbox{\ion{Si}{4}})$ for the $\rho$ Leo sightline only applies to
the velocity range $v_\odot = -19$ to $+2$ \kms\ (the range over which
the \ion{S}{3} measurements are made).  There is detectable
\protect\ion{Si}{4} absorption at more positive velocities.  The integrated
sightline column density is $\log N(\mbox{\protect\ion{Si}{4}}) =
12.24^{+0.07}_{-0.09}$.}
\end{planotable}

\pagebreak

\pagebreak

\begin{references}
\reference{} Aannestad, P.A., 1989, \apj, 338, 162

\reference{} Ali, B., Blum, R.D., Bumgardner, T.E., Cranmer, S.R.,
Ferland, G.J., Haefner, R.I., \& Tiede, G.P., 1991, \pasp, 103, 1182

\reference{} Anders, E., \& Grevesse, N. 1989,
Geochem. Cosmochem. Acta, 53, 197

\reference{} Baldwin, J.A. \etal, 1996, \apj, 468, L115

\reference{} Barker, E.S., Lugger, P.M., Weiler, E.J., \& York,
D.G. 1984, \apj, 280, 600

\reference{} Bautista, M.A. \& Pradhan, A.K. 1995, \apj, 442, L65

\reference{} Bautista, M.A. \& Pradhan, A.K. 1998, \apj, 492, 650

\reference{} Bohlin, R.C., Savage, B.D., \& Drake, J.F. 1978, \apj,
224, 132

\reference{} Boulanger, F., Abergel, A., Bernard, J.-P., Burton, W.B.,
D\'{e}sert, F.-X., Hartmann, D., Lagache, G., \& Puget, J.-L. 1996,
\aap, 312, 256

\reference{} Brandt, \etal\ 1998, \aj, submitted.

\reference{} Cardelli, J.A., Ebbets, D.C., \& Savage, B.D. 1993, \apj,
413, 401

\reference{} Cardelli, J.A., \& Ebbets, D.C. 1994, in {\em HST}
Calibration Workshop, Calibrating {\em Hubble Space Telescope},
ed. J.C. Blades \& A.J. Osmer (Baltimore: STScI), 322

\reference{} Cardelli, J.A., Meyer, D.M., Jura, M., \& Savage,
B.D. 1996, \apj, 467, 334

\reference{} Code, A.D., Bless, R.C., Davis, J., \& Brown, R.H. 1976,
\apj, 203, 417

\reference{} Dettmar, R.-J., \& Schulz, H. 1992, \aap, 254, L25

\reference{} Dickey, J.M., \& Lockman, F.J., 1990, \araa, 28, 215

\reference{} Diplas, A., \& Savage, B.D. 1994, \apjs, 93, 211

\reference{} Domg\"{o}rgen, H. \& Mathis, J.S. 1994, \apj, 428, 647 (DM94)

\reference{} Duncan, D.K. 1992, \ghrs\ Handbook, Version 3.0, Space
Telescope Science Institute

\reference{} Edgar, R.J. \& Savage, B.D., 1989, \apj, 340, 762

\reference{} Edvardsson, B., Andersen, J., Gustafsson, B., Lambert,
D.L., Nissen, P.E., \& Tomkin, J. 1993, \aap, 275, 101

\reference{} Federman, S.R., Sheffer, Y., Lambert, D.L., \& Gilliland,
R.L. 1993, \apj, 413, L51

\reference{} Ferland, G.J. 1996, Hazy, a Brief Introduction to CLOUDY
 90, University of Kentucky Department of Physics and Astronomy
 Internal Report

\reference{} Ferland, G.J., Korista, K.T., Verner, D.A., Ferguson,
J.W., Kingdon, J.B., \& Verner, E.M. 1998, \pasp, 110, 761

\reference{} Fitzpatrick, E.L., \& Spitzer, L. 1994, \apj, 427, 258

\reference{} Fitzpatrick, E.L., \& Spitzer, L. 1997, \apj, 475, 623


\reference{} Grevesse, N., \& Noels, A. 1993, in Origin of the
Elements, ed., N. Prantzos, E. Vangioni-Flam, M. Casse\'{e},
(Cambridge: Cambridge Univ. Press), p. 15

\reference{} Gry, C., Lequeux, J., \& Boulanger, F. 1992, \aap, 266, 457

\reference{} Heap, S.R. \etal\ 1995, \pasp, 107, 871

\reference{} Henry, R.B.C. 1993, \mnras, 261, 306

\reference{} Holmgren, D., Hadrava, P., Harmanec, P., Koubsky, P., \&
Kubat, J. 1997, \aap, 322, 565

\reference{} Howarth, I.D., \& Prinja, R.K. 1989, \apjs, 69, 527

\reference{} Howk, J.C., Savage, B.D., \& Fabian, D. 1998, in preparation.

\reference{} Jenkins, E.B. 1983, in Kinematics, Dynamics and Structure
of the Milky Way, ed. W.L.H. Shuter (Dordrecht: Reidel), p. 21.

\reference{} Jenkins, E.B. 1987, in Interstellar Processes, ed.
D.J. Hollenbach \& H.A. Thronson, Jr. (Dordrecht: Reidel), p. 533.

\reference{} Jenkins, E.B., Savage, B.D., \& Spitzer, L. 1986, \apj,
301, 355

\reference{} Keenan, F.P., Brown, P.J.F., \& Lennon, D.J. 1986, \aap,
155, 333

\reference{} Keenan, F.P., \& Dufton, P.L. 1983, \mnras, 205, 435

\reference{} Kulkarni, S.R., \& Heiles, C. 1988, in Galactic and
Extragalactic Radio Astronomy, ed. K.I. Kellerman, G.L. Verschuur,
(New York: Springer-Verlag), p. 95

\reference{} Kurucz, R.L. 1991, in Proceedings of the Workshop on
Precision Photometry: Astrophysics of the Galaxy, ed. A.C. Davis
Philip, A.R. Upgren, \& K.A. James (Schenectady: Davis), p. 27.

\reference{} Mathis, J.S. 1986a, \apj, 301, 423

\reference{} Mathis, J.S. 1986b, \pasp, 98, 995


\reference{} McGaugh, S.S. 1991, \apj, 380, 140

\reference{} McKee, C.F. \& Ostriker, J.P. 1977, \apj, 218, 148

\reference{} Morton, D.C. 1975, \apj, 197, 85

\reference{} Morton, D.C. 1991, ApJS, 77, 119

\reference{} Nussbaumer, H., \& Storey, P.J. 1983, \aap, 126, 75

\reference{} Nussbaumer, H., \& Storey, P.J. 1986, \aaps, 64, 545

\reference{} Osterbrock, D.E. 1989, Astrophysics of Gaseous Nebulae
and Active Galactic Nuclei (Mill Valley: University Science Books)

\reference{} Osterbrock, D.E., Tran, H.D., \& Veilleux, S. 1992, \apj, 389

\reference{} Peimbert, M., \& Goldsmith, D.W. 1972, \aap, 19, 398

\reference{} Peimbert, M., Torres-Peimbert, S., \& Dufour, R.J. 1993,
\apj, 418, 760
 
\reference{} Perryman, M.A.C. an dthe Hipparcos Science Team 1997, The
Hipparcos and Tycho Catalogues, ESA SP 1200, (Noordwijk: ESA Publications).

\reference{} Rand, R.J. 1997, \apj, 474, 129

\reference{} Rand, R.J. 1998, \apj, 501, 137

\reference{} Reynolds, R.J. 1984, \apj, 282, 191

\reference{} Reynolds, R.J. 1985, \apj, 294, 256


\reference{} Reynolds, R.J. 1988b, \apj, 333, 341

\reference{} Reynolds, R.J. 1989, \apj, 339, L29

\reference{} Reynolds, R.J. 1991a, \apj, 372, L17

\reference{} Reynolds, R.J. 1991b, in The Interstellar Disk-Halo
Connection in Galaxies; IAU Symp. No. 144, ed. H. Bloemen
(Dordrecht: Kluwer), p. 67.

\reference{} Reynolds, R.J., \& Cox, D.P. 1992, 400, L33

\reference{} Reynolds, R.J., Hausen, N.R., Tufte, S.L., \& Haffner,
L.M. 1998a, \apj, 494, L99

\reference{} Reynolds, R.J., \& Ogden, P.M. 1982, \aj, 87, 306

\reference{} Reynolds, R.J., \& Tufte, S.L. 1995, 439, L17

\reference{} Reynolds, R.J., Tufte, S.L., Haffner, L.M., Jaehnig, K.,
\& Percival, J.W. 1998b, PASA, 15, 14


\reference{} Robinson, R.D. \etal\ 1998, \pasp, 110, 68

\reference{} Rodr\'{\i}guez, M. 1996 \aap, 313, L5

\reference{} Rubin, R.H., Dufour, R.J., Ferland, G.J., Martin, P.G.,
O'Dell, C.R., Baldwin, J.A., Hester, J.J., Walter, D.K., \& Wen,
Z. 1997, \apj, 474, L131

\reference{} Ryans, R.S.I., Sembach, K.R., \& Keenan, F.P. 1996, \aap,
314, 609

\reference{} Savage, B.D., \& Bohlin, R.C. 1979, \apj, 229, 136

\reference{} Savage, B.D., Cardelli, J.A., \& Sofia, U. J. 1992, \apj,
401, 706

\reference{} Savage, B.D., Edgar, R.J., \& Diplas, A. 1990, \apj, 361, 107


\reference{} Savage, B.D., \& Sembach, K.R. 1991, \apj, 379, 245

\reference{} Savage, B.D., \& Sembach, K.R. 1994, \apj, 434, 145

\reference{} Savage, B.D., \& Sembach, K.R. 1996a, \apj, 470, 893

\reference{} Savage, B.D., \& Sembach, K.R. 1996b, \araa, 34, 279

\reference{} Sciama, D.W. 1995, \apj, 448, 667

\reference{} Sciama, D.W. 1997, \apj, 488, 234

\reference{} Sembach, K.R., \& Savage, B.D. 1992, \apjs, 83, 147

\reference{} Sembach, K.R., \& Savage, B.D. 1994, \apj, 434, 145

\reference{} Sembach, K.R., Savage, B.D., \& Jenkins, E.B.  1994,
\apj, 421, 585

\reference{} Shields, J.C., \& Kennicutt, R.C. 1995, \apj, 454, 807

\reference{} Shull, J.M., \& York, D.G. 1977, \apj, 211, 803

\reference{} Soderblom, D.R., Sherbert, L.E., \& Hulbert, S.J. 1993,
GHRS Instrument Science Report 52 (Baltimore: STScI)

\reference{} Soderblom, D.R., Sherbert, L.E., \& Hulbert, S.J. 1994,
GHRS Instrument Science Report 53 (Baltimore: STScI)

\reference{} Sofia, U.J., \& Jenkins, E.B., 1998, \apj, in press.


\reference{} Sofia, U.J., Savage, B.D., \& Cardelli, J.A. 1993, \apj,
413, 251

\reference{} Sokolov, N.A. 1995, \aaps, 110, 553

\reference{} Spitzer, L. 1985, \apj, 290, L21

\reference{} Spitzer, L., \& Fitzpatrick, E.L. 1993, \apj, 409, 299
(SF93)

\reference{} Spitzer, L., \& Fitzpatrick, E.L. 1995, \apj, 445, 196

\reference{} Sutherland, R.S., \& Dopita, M.A. 1993, \apjs, 88, 253



\reference{} Taylor, J.H., \& Cordes, J.M. 1993, \apj, 411, 674

\reference{} Vacca, W.D., Garmany, C.D., \& Shull, J.M., 1996, \apj,
460, 914

\reference{} Verner, D.A., Ferland, G., Korista, K., \& Yakovlev,
D.G., 1996, \apj, 465, 487

\reference{} Wheeler, J.C., Sneden, C., \& Truran, J.W. 1989, \araa,
27, 279

\reference{} Wolfire, M.G., Hollenbach, D., McKee, C.F., Tielens,
A.G.G.M., \& Bakes, E.L.O. 1995, \apj, 443, 152

\reference{} Wolfire, M.G., McKee, C.F.,  Hollenbach, D., \& Tielens,
A.G.G.M. 1995, \apj, 453, 673



\end{references}
\end{document}